\documentclass[fleqn,usenatbib,useAMS]{mnras}

\usepackage{graphicx}	
\usepackage{amsmath}	
\usepackage{amssymb}	
\usepackage{multicol}        
\usepackage{bm}		
\usepackage{pdflscape}	

\usepackage[T1]{fontenc}
\usepackage{ae,aecompl}

\usepackage{newtxtext,newtxmath}

\title{Exploring the ex-situ components within $Gaia$ DR3}

\author{Zhuohan Li$^{1,2}$}
\author[Li et al.]{
Zhuohan Li$^{1,2}$,
Gang Zhao$^{1,2}$, \thanks{Contact e-mail: \href{gzhao@nao.cas.cn}{gzhao@nao.cas.cn}}
Ruizhi Zhang$^{1,2}$,
Xiang-Xiang Xue$^{1,3}$,
Yuqin Chen$^{1,2}$,
João A. S. Amarante$^{4,5}$\thanks{Visiting Fellow at UCLan}
\\
$^{1}$ CAS Key Laboratory of Optical Astronomy, National Astronomical Observatories, Chinese Academy of Sciences, Beijing 100101, People's Republic of China
\\
$^{2}$ School of Astronomy and Space Science, University of Chinese Academy of Sciences, Beijing 100049, People’s Republic of China
\\
$^{3}$ Institute for Frontiers in Astronomy and Astrophysics, Beijing Normal University, Beijing, 102206, People’s Republic of China
\\
$^{4}$Institut de Ciencies del Cosmos (ICCUB), Universitat de Barcelona (IEEC-UB), Martí i Franquès 1, E08028 Barcelona, Spain
\\
$^{5}$Jeremiah Horrocks Institute, University of Central Lancashire, Preston, PR1 2HE, UK}

\date{Last updated 2020 June 10; in original form 2013 September 5}

\pubyear{2020}

\begin{document}
\label{firstpage}
\pagerange{\pageref{firstpage}--\pageref{lastpage}}
\maketitle

\begin{abstract}
The presence of $Gaia$ DR3 provides a large sample of stars with complete 6D information, offering a fertile ground for the exploration of stellar objects that were accreted to the Milky Way through ancient merger events. In this study, we developed a deep learning methodology to identify ex-situ stars within the $Gaia$ DR3 catalogue. After two phases of training, our neural network (NN) model was capable of performing binary classification of stars based on input data consisting of 3D position and velocity, as well as actions. From the target sample of 27 085 748 stars, our NN model managed to identify 160 146 ex-situ stars. The metallicity distribution suggests that this ex-situ sample comprises multiple components but appears to be predominated by the Gaia-Sausage-Enceladus. We identified member stars of the Magellanic Clouds, Sagittarius, and 20 globular clusters throughout our examination. Furthermore, an extensive group of member stars from Gaia-Sausage-Enceladus, Thamnos, Sequoia, Helmi streams, Wukong, and Pontus were meticulously selected, constituting an ideal sample for the comprehensive study of substructures. Finally, we conducted a preliminary estimation to determine the proportions of ex-situ stars in the thin disc, thick disc, and halo, which resulted in percentages of 0.1\%, 1.6\%, and 63.2\%, respectively. As the vertical height from the Galactic disc and distance from the Galactic centre increased, there was a corresponding upward trend in the ex-situ fraction of the target sample.
\end{abstract}




\begingroup
\let\clearpage\relax
\endgroup
\newpage

\section{Introduction}
The hierarchical merger hypothesis \citep{white1978core} has long been a prominent theory in explaining the formation of galaxies. Within the framework of the Lambda Cold Dark Matter ($\Lambda$CDM) cosmology and in light of current observations, a picture has been constructed in which large galaxies are formed through a process of hierarchical mergers with smaller satellite galaxies. For instance, the ongoing merger between the Milky Way (MW) and the Sagittarius dwarf spheroidal galaxy (Sgr; \citealt{ibata1994dwarf}) provides direct observational evidence of the anticipated accretion events. Furthermore, remnants of ancient accretion events have also been observed in the form of kinematic substructures and stellar streams (e.g. \citealt{zhao2009catalog, xue2011quantifying, zhao2015halo, helmi2020streams}).\par
Facilitated by the availability of high-quality astrometric data for a large number of stars, the $Gaia$ mission \citep{TheGaiamission} has initiated a new era in the investigation of Galactic formation. One of the most notable substructures confirmed through studies based on $Gaia$ data \citep{GaiaDR1, gaia2018gaia} is the Gaia-Sausage-Enceladus (GSE; \citealt{belokurov2018co, helmi2018merger, myeong2018sausage, haywood2018disguise}). It is believed to be the remnant of the last major merger event experienced by the MW, which was completed approximately 10 Gyr ago \citep{gallart2019uncovering, helmi2020streams}. Kinematic studies of other substructures, including Sequoia \citep{myeong2019evidence}, Thamnos \citep{Koppelman2019T} and the Helmi streams \citep{helmi1999debris, HS2019} were also conducted contemporaneously.
\par
On the other hand, spectroscopic surveys such as the RAdial Velocity Experiment (RAVE; \citealt{steinmetz2006radial}), Sloan Extension for Galactic Understanding and Exploration (SEGUE; \citealt{yanny2009segue}), LAMOST survey \citep{zhao2006stellar, zhao2012lamost, deng2012lamost, liu2013lss, luo2015first, yan2022overview}, Galactic Archaeology with HERMES (GALAH; \citealt{de2015galah}), Apache Point Observatory Galactic Evolution Experiment (APOGEE; \citealt{majewski2017apache}) and H3 survey \citep{conroy2019mapping}, provide us with insights into chemical abundances. The formation and chemical enrichment history of dwarf galaxies differs from that of the MW, and dwarf galaxies of different masses exhibit distinct chemical evolution patterns. Based on chemical abundance data obtained from the H3 survey \citep{conroy2019mapping}, \citet{naidu2020evidence} identified both previously known and newly discovered structures, such as Aleph, Arjuna, I'itoi and Wukong. Another chemo-dynamical study that employed data from the APOGEE DR16 \citep{ahumada202016th} revealed evidence for the existence of a metal-poor stellar structure in the inner Galaxy \citep{horta2021evidence}, which was previously identified as the Kraken \citep{kruijssen2020kraken} and may be a
signature of the proto-MW \citep{Belokurov2022Aurora, rix2022poor}. With the support of APOGEE DR17 \citep{accetta2022seventeenth}, \citet{horta2022chemical} further conducted a comprehensive investigation into the chemical patterns exhibited by various previously known substructures and provided an in-depth analysis of their origins. Moreover, chemically peculiar stars were examined in APOGEE data (e.g. \citealt{fernandez2019chemodynamics, fernandez2022Galactic}), offering further insights into the merger and evolutionary history of the MW.
\par
The incorporation of chemical abundance significantly enhances the analysis of substructures. However, there is a substantial discrepancy in sample sizes between spectroscopic and astrometric surveys. The third data release of $Gaia$ \citep{GaiaDR3} provided us with astrometric data for over 1.8 billion stars, of which more than 30 million have 6D phase space information. Concurrently, the number of sources observed by a single high-resolution spectroscopic survey is limited to the order of one million \citep{buder2021galah+, accetta2022seventeenth}. The LAMOST survey has achieved tens of millions of spectra, leading to numerous significant findings (e.g. \citealt{li2015spectroscopic, li2018catalog, yan2018nature, yan2021most, xing2019evidence, xing2023metal, zhao2021low}). However, as a median-to-low resolution survey, the preponderance of its data is confined to information on metal abundance and a limited number of elements. Given the current circumstances, it would be highly beneficial for the community to devise a data-driven methodology that solely relies on kinematic data to attain comparable selection outcomes as those obtained through the utilization of multi-element abundance information.
\par
Data-driven approaches have already been utilized in researches endeavor to detect merger debris within the MW. In the pursuit of uncovering the aforementioned substructures, several classic machine learning algorithms have been employed, including KNN \citep{fix1951discriminatory}, GMM \citep{dempster1977maximum} and HDBSCAN \citep{campello2013density}. Recent efforts to identify stellar streams and kinematic substructures have also incorporated the use of machine learning algorithms. By performing clustering using the DBSCAN algorithm \citep{ester1996dbscan} within a 6D phase space, \citet{borsato2020identifying} were able to detect five high-confidence streams, including one that had not been previously discovered. \citet{yuan2020dynamical} employed a self-organizing map algorithm \citep{kohonen2001self, yuan2018stargo} within a 4D space of orbital energy and angular momentum to identify 57 dynamically tagged groups, most of which belong to previously known substructures such as the GSE. Building upon the ANODE algorithm \citep{nachman2020anomaly}, \citet{shih2022via} developed a novel methodology, termed the VIA MACHINE, for the identification of cold stellar streams within data acquired from the $Gaia$. Through the implementation of this approach, they were able to successfully identify the presence of the GD-1 stream \citep{grillmair2006detection}. By conducting t-Distributed Stochastic Neighbor Embedding (t-SNE; \citealt{van2008visualizing}) analysis in chemical space, \cite{ortigoza2023Galactic} identified seven structures, including the Splash, GSE, the high-$\alpha$ heated-disc population, N-C-O peculiar stars, inner disc-like stars, and two previously unreported structures. Unsupervised learning algorithms do not necessitate prior labeling of data. Instead, the algorithms make decisions based on the inherent distribution of the data. Nevertheless, the absence of a labeling process makes it challenging to artificially control the behavior of unsupervised algorithms and to quantify their performance. Consequently, it cannot be guaranteed that structures with clustering properties in phase space are related to ancient merger events. These structures may also originate from the disruption of local globular clusters in the MW. Therefore, it is important to ensure a pure ex-situ sample before using clustering algorithms for detailed analysis.
\par
Deep learning allows computational models that are composed of multiple processing layers to learn representations of data with multiple levels of abstraction \citep{lecun2015deep}, it holds considerable promise as a tool in the search for ex-situ components. Through the implementation of supervised deep learning techniques on a known dataset, an optimal model can be derived for the purpose of classification. \citet{ostdiek2020cataloging} implemented an NN trained to identify accreted stars using only 5D kinematics information and construct a catalogue comprising 767 000 accreted stars. Subsequent companion works revealed the existence of a vast prograde stellar stream in the solar vicinity, which they named Nyx \citep{necib2020evidence, necib2020chasing}. The methodology presented by \citet{ostdiek2020cataloging} provides an instructive paradigm for the application of supervised learning to the identification of merger debris. By pre-training the model on simulated data, it is able to classify accreted stars labeled at truth level and capture their major kinematic features. The employment of transfer learning further improved the model’s ability to adapt to real data of the MW. However, the absence of radial velocity limits the classification to kinematic level, as the derivation of integrals of motion requires full 6D phase space information. With the availability of $Gaia$ DR3 data and the inclusion of photo-astrometric distances derived by the \texttt{StarHorse} code \citep{queiroz2018starhorse, anders2022photo}, the sample size of stars with accurately determined full 6D phase space parameters is no longer a constraining factor. In addition, recent data releases of spectroscopic surveys have facilitated the tagging of accreted stars in a more efficient manner, thereby enabling the improvement of the second phase of network training based on previous research.\par
In this study, we present a deep learning methodology for the identification of ex-situ stars through the utilization of 6D phase space parameters and the actions. Following the completion of two training phases, our NN model demonstrated the ability to effectively identify ex-situ components comprised of dwarf galaxies, globular clusters, and merger debris within the target sample of $Gaia$ DR3. We conducted a preliminary analysis of the NN classified sample and dedicate it to the community in the hope that it could provide novel insights and inspiration. Further analysis, including clustering and chemical follow-up, may be performed to fully explore the potential of this sample and deepen our understanding of the Milky Way's merger history.
\par
This paper is organized as follows: Section~\ref{section:sec2} introduces the NN model, the training process, and the data utilized in this study. In Section~\ref{section:sec3}, we present the classification results of the NN and provide an initial analysis of the ex-situ sample. Section~\ref{section:sec4} discusses particular elements involved in our algorithm and analysis. Finally, we summarize the results of this study in Section~\ref{section:sec5}.

\section{Method}\label{section:sec2}
\subsection{Base model}
As in \citet{ostdiek2020cataloging}, we constructed an NN model that was trained on synthetic data to learn fundamental kinematic features. The NN model was developed and trained using the \texttt{Keras} \citep{chollet2015keras} library, with \texttt{TensorFlow} \citep{tensorflow2015-whitepaper} serving as the backend framework. The input to the NN consists of three-dimensional position and velocity data in the Cartesian coordinate system. The model comprises five fully connected hidden layers, with the intermediate three layers containing 128 nodes and the remaining two layers containing 64 nodes. We selected the Rectified Linear Unit (ReLU; \citealt{hahnloser2000digital}) as the activation function, and applied batch normalization prior to each activation. The output layer of the NN is composed of a single node that uses a sigmoid activation function. This function scales the output to fall within the range of 0 to 1, where values closer to 1 indicate a higher probability of the star being ex-situ. We refer to this NN model as \texttt{NN\_FIRE} in subsequent discussions.

\subsubsection {Synthetic data}\label{section:sec2.1.1}
As a supervised learning algorithm, the performance of the NN is greatly influenced by the labels of the training set. To obtain truth level ex-situ labels, we train our NN with simulated data, where the accretion history of each star is traceable. The state-of-the-art cosmological hydrodynamic simulations, specifically FIRE-2 \citep{hopkins2018fire, wetzel2023public}, provide us a robust platform for our training endeavors. Rather than directly employing the raw data of FIRE-2 simulations, we opted to utilize a mock $Gaia$ catalogue constructed by \citet{sanderson2020synthetic} to facilitate the alignment between the patterns of the simulated data and the observational data from $Gaia$ satellite.\par
Detailed elaboration of the synthetic catalogue is presented in \citet{sanderson2020synthetic}, and reviewed by \citet{ostdiek2020cataloging}. The mock catalogues are derived from three MW-mass galaxies from the $Latte$ suite of FIRE-2 simulations, namely \texttt{m12i}, \texttt{m12f} and \texttt{m12m}. Assuming each star particle of mass $\approx 7000 M_{\odot}$ represents a single stellar population, \citet{sanderson2020synthetic} sampled synthetic stars following the algorithm described in \citet{sharma2011galaxia}. Three different local standard of rest (LSR) were deployed in each simulated galaxy as "solar viewpoints" to construct the synthetic catalogue. The data structure of the synthetic catalogue resembles that of the $Gaia$ catalogue, which supplies astrometric and photometric data. As in \citet{ostdiek2020cataloging}, we utilized data from all three LSRs in the \texttt{m12i}, encompassing a total of 47 673 267 synthetic stars. We only selected stars with available error-convolved radial velocity, which covers a distance range of 0 to 3 kpc. The data utilized included error-convolved right ascension (\texttt{ra}), declination (\texttt{dec}), proper motion in right ascension (\texttt{pmra}), proper motion in declination (\texttt{pmdec}) and radial velocity (\texttt{radial\_velocity}). In addition, the true LSR-centric distance (\texttt{dhel\_true}) was used instead of the error-convolved parallax (\texttt{parallax}) to ensure compatibility with the final model. The astrometric data were then transformed into 3D position and velocity in the Galactocentric Cartesian coordinate system using the \texttt{astropy} package \citep{price2018astropy}, as the input of the NN.
\par
Although the synthetic catalogue is build to mock $Gaia$ DR2, we note that it is not necessary to reproduce the exact selection function of the latest data release of $Gaia$. The primary objective of \texttt{NN\_FIRE} is to acquire fundamental knowledge of stellar kinematics. The inconsistency between the distribution of training samples and target samples will be resolved in the subsequent training process, as described in Section~\ref{section:sec2.2}.\par
To identify the ex-situ stars in the synthetic catalogue, we made use of a text file named \texttt{star\_exsitu\_flag\_600.txt} in the \texttt{m12i} directory \citep{wetzel2023public}, which provides a binary flag for each star particle at redshift equals to 0. A flag value of 1 indicates that the star particle is ex-situ, while a value of 0 indicates that it is in-situ. Star particles are classified as “ex-situ” if they originated at a spherical distance greater than 30 kpc comoving from the centre of the primary galaxy \citep{bellardini20223d}. Synthetic stars generated from the same star particle are assigned the same \texttt{parentid}, which is the array index of the star particle. We labeled each star with the binary flag corresponding to the index in the text file, resulting in a total of 380 326 ex-situ stars.

\subsubsection {Training and evaluation}\label{section:sec2.1.2}
Our training process adheres to the standard workflow within the \texttt{Keras} framework. We employed the \texttt{Adam} optimizer \citep{kingma2014adam}, with an initial learning rate of $10^{-3}$.  A learning rate reduction mechanism was implemented, whereby if the validation loss failed to decrease for 5 consecutive epochs, the learning rate would be automatically reduced to half of its original value. The minimum learning rate threshold was set at $10^{-6}$. Furthermore, to avoid overfitting, the training process would be terminated if the validation loss failed to decrease for 10 epochs. However, this early stopping mechanism was not triggered during a total of 50 epochs of training.
\par
The synthetic stars in our dataset were partitioned into training, validation, and test sets at a ratio of 6:2:2. The 6D phase space data in these three datasets were normalized using a \texttt{StandardScaler} from the \texttt{sklearn.preprocessing} module \citep{pedregosa2011scikit}, which was fit to the training set.  The partitioning process was accomplished using the \texttt{sklearn.train\_test\_split} function, during which we employed stratified sampling on the label to ensure that the proportion of ex-situ stars remained consistent across all datasets. However, ex-situ stars comprise a very small fraction of our synthetic sample, which presents a classification problem with an extreme class imbalance. One common method to address class imbalance is sample-weighted training, but determining the appropriate weight can be challenging. If the weight is set too high, it can skew the distribution of the prediction and result in poor classification performance. Conversely, if the weight is set too low, it may not effectively address the issue of imbalanced sample size. Therefore, we aim to solve this problem by choosing a proper loss function. The \texttt{Focal Loss} is a loss function introduced by \citet{lin2017focal} to mitigate the issue of class imbalance during training in tasks such as binary classification. This function applies a modulating term to the cross-entropy loss to focus on the samples that are challenging to classify. The function is mathematically defined as:
\begin{equation}
FL(p)=
\begin{cases}
-(1-p)^{\gamma}\log(p) & \text{if y = 1} \\
-p^{\gamma}\log(1-p) & \text{if y = 0}
\end{cases}
\end{equation}
where p denotes the output of the classifier and $\gamma$ represents a tunable hyper parameter. In our training process, we adopted the default setting of $\gamma$=2. As the confidence in the correct class increases, the scaling factor decays to zero, automatically down-weighting the contribution of samples that are easier to classify during training and rapidly focusing the model on more challenging samples.\par
Due to the extreme class imbalance in our classification problem, the commonly used metric of accuracy is not a suitable measure of NN classification performance. Accuracy reflects the proportion of samples that are correctly classified within the entire sample. However, in our sample, less than 1\% of the stars are labeled as ex-situ. As a result, a classifier that indiscriminately assigns all stars as in-situ could easily achieve an accuracy rate exceeding 99\%, even in the absence of any classification activity.\par
In order to adopt a more appropriate metric, we initially consider ex-situ stars as positive samples while in-situ stars the negative samples. With respect to the classification results, ex-situ samples that are accurately identified by the NN are defined as true positive samples (TP). In contrast, ex-situ samples that are erroneously classified as in-situ by the NN are defined as false negative samples (FN). Correspondingly, in-situ samples that are accurately identified by the NN are regarded as true negative samples (TN), while in-situ samples that are erroneously classified as ex-situ by the NN are regarded as false positive samples (FP). To evaluate the purity of ex-situ samples selected by the NN, we employed the metric of precision:
\begin{equation}
Precision=
\rm
\frac{TP}{TP+FP}
\end{equation}
Furthermore, the true positive rate (TPR), commonly referred to as recall, illustrates the completeness of the selection:
\begin{equation}
TPR=
\rm
\frac{TP}{TP+FN}
\end{equation}
Additionally, we calculated the false positive rate (FPR) to quantify the proportion of in-situ samples that were incorrectly classified as ex-situ:
\begin{equation}
FPR=
\rm
\frac{FP}{TN+FP}
\end{equation}
Precision and recall are often considered to be conflicting measures, making it challenging for a classifier to achieve high values for both simultaneously. In the task of identifying ex-situ stars, our goal is to ensure that the ex-situ samples selected by the NN have a high degree of purity, which corresponds to a higher precision. Similarly, we also strive to minimize the FPR. However, this may result in sacrificing the completeness of the ex-situ sample, leading to a relatively lower recall.
\par
\begin{figure}
 \includegraphics[width=\columnwidth]{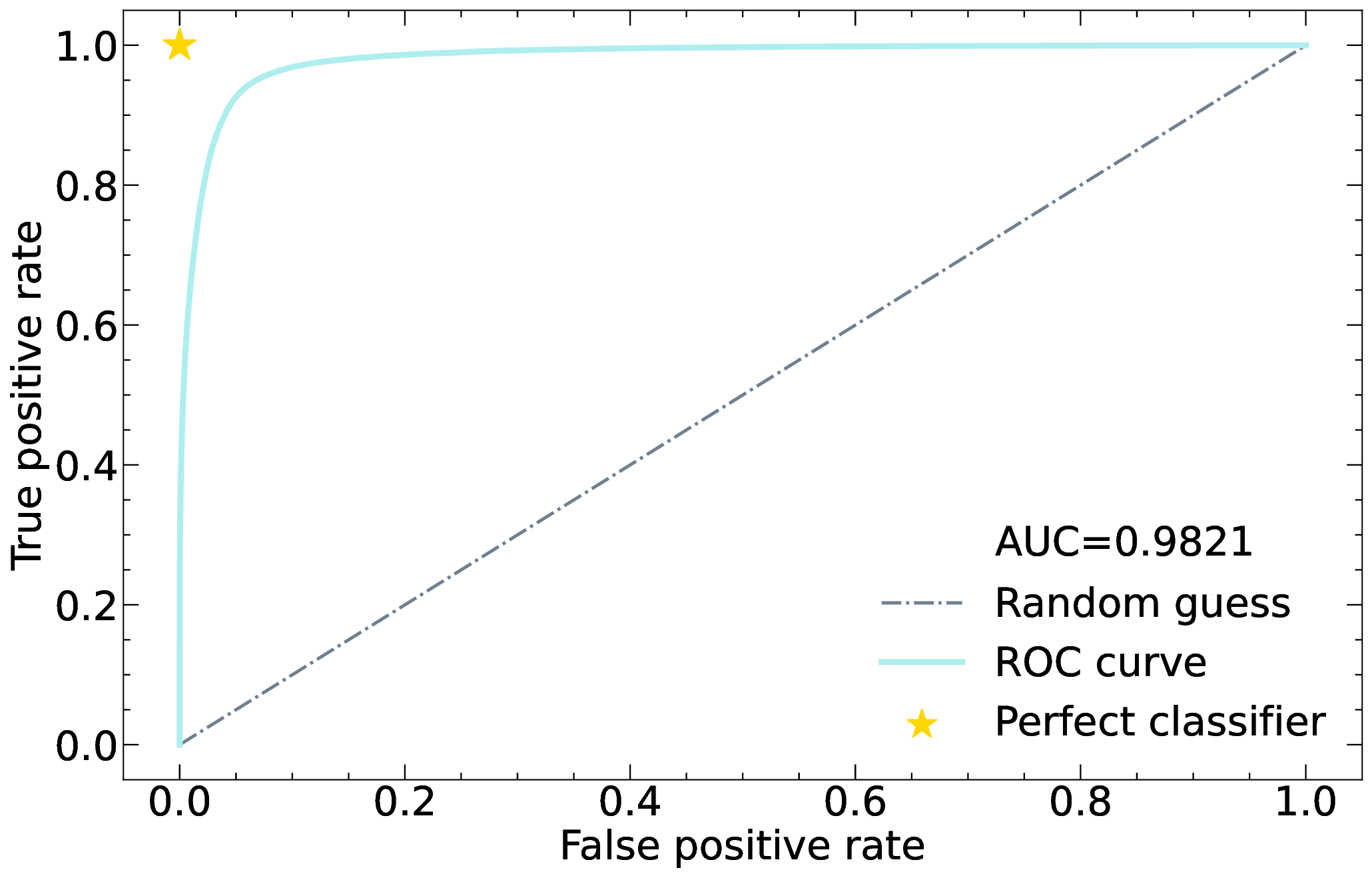}
 \caption{As demonstrated by the ROC curve, the \texttt{NN\_FIRE} achieves an AUC of 0.98 on the test set. The performance of a perfect classifier, indicated by a gold star, has a TPR of 1 and an FPR of 0. The grey dashed line represents the performance of a random guess.}
 \label{fig:ROC_curve}
\end{figure}

The output of an NN consists of a series of continuous values. In order to map these values to discrete labels, it is necessary to establish a threshold. The final classification result can be expressed as:
\begin{equation}
Prediction =
\begin{cases}
1 & \text{if } p > \text{threshold} \\
0 & \text{if } p \leq \text{threshold}
\end{cases}
\end{equation}
where p represents the raw output of the NN. The choice of threshold can have a significant impact on the performance of the classifier, as different thresholds will result in varying levels of TPR and FPR. By selecting a series of thresholds, a corresponding series of TPRs and FPRs can be obtained, allowing for the construction of a Receiver Operating Characteristic (ROC) curve. According to the performance of \texttt{NN\_FIRE} on the test set, we present the ROC curve as shown in Figure~\ref{fig:ROC_curve}. The Area Under the Curve (AUC) is a quantitative measure of classifier performance, with values ranging from 0 to 1. An AUC value closer to 1 indicates superior classifier performance. In this case, \texttt{NN\_FIRE} obtained an AUC value exceeding 0.98, denoting an excellent result. As represented by the gold star, an ideal classifier would achieve a TPR of 1 and a FPR of 0, corresponding to a precision of 100\%. However, as previously discussed, this is not achievable in practice and a trade-off must be made between the purity and completeness of the ex-situ sample. When the threshold was established at 0.5, \texttt{NN\_FIRE} had a precision of 80.1\% and a recall of 34.5\%. Elevating the threshold to 0.75 yielded a precision of 98.3\%, but diminished the recall to 13.4\%. As \texttt{NN\_FIRE} is not the final model, we will refrain from discussing the threshold setting in detail at present. Based on the analysis of the ROC curve, we consider that the initial training phase has imbued \texttt{NN\_FIRE} with the ability to identify ex-situ stars from a kinematic perspective.

\subsection{Parallel model}\label{section:sec2.2}
In the context of NN regression tasks, the distribution of labels within the training set warrants substantial attention. Predictions outside this range or in sparse parts of the distribution are regarded as extrapolations of the model. Unlike regression tasks where label values are continuous, our classification task has only two categories, corresponding to label values 0 and 1. In this case, caution should still be exercised regarding extrapolation, but the focus should be on the distribution of NN input parameters. During the training phase of \texttt{NN\_FIRE}, only sources with distances less than 3 kpc are considered. However, the target sample may encompass distances reaching tens of kiloparsecs. Moreover, even if the simulated data is derived from a synthetic $Gaia$ survey catalogue, it may not precisely align with observational data. Therefore, we need to design a training framework to improve our model’s performance on the target sample.
\par
\begin{figure}
 \includegraphics[width=\columnwidth]{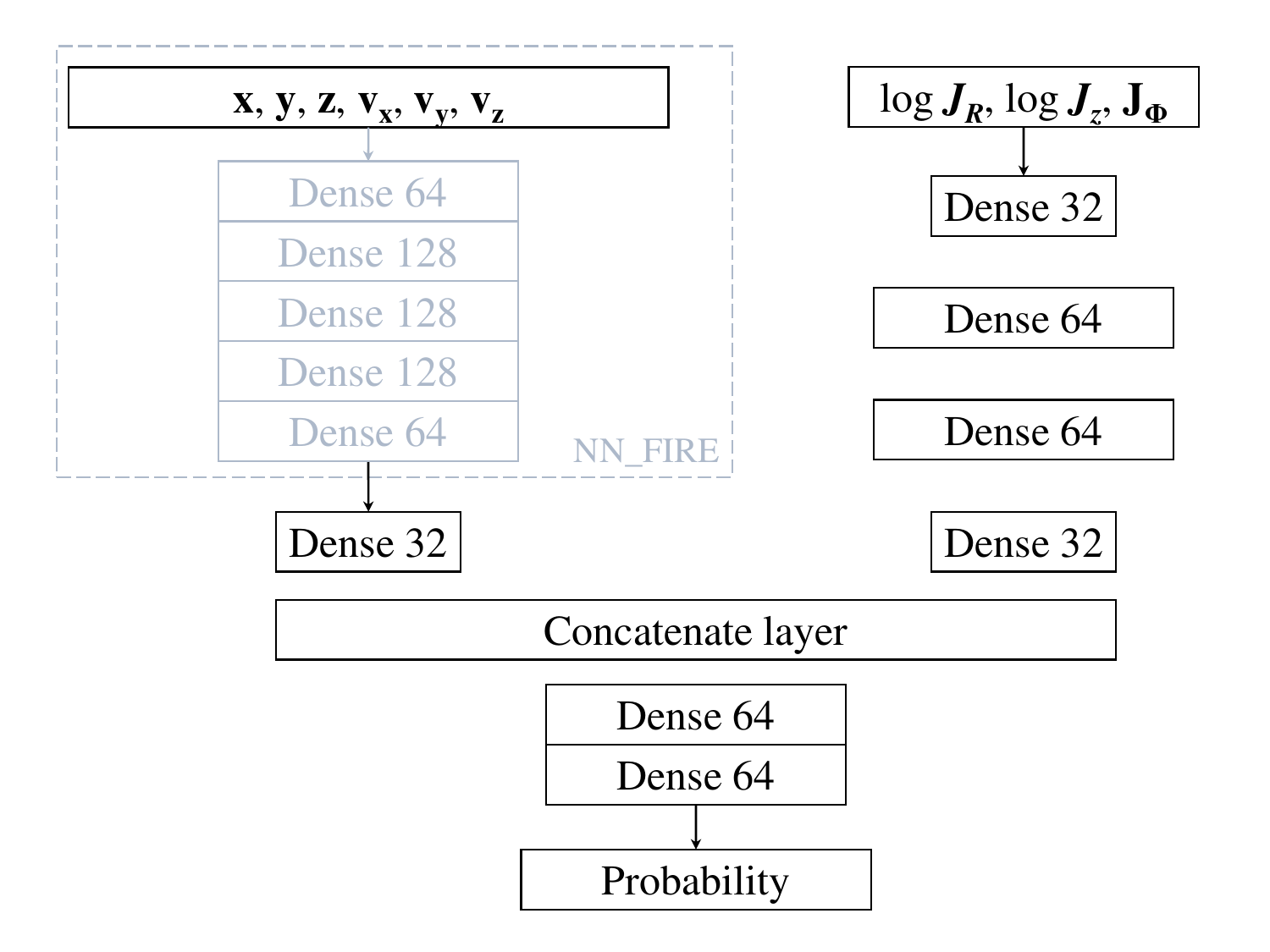}
 \caption{The parallel model comprises two components: the \texttt{NN\_FIRE} network and a sub-network that processes actions. The outputs of these two components are concatenated and subsequently processed by a series of dense layers. The weighted layers of the \texttt{NN\_FIRE} network are frozen, as indicated by their grey coloration.}
 \label{fig:NN_parallel}
\end{figure}

Instead of performing the standard workflow of transfer learning as in \citet{ostdiek2020cataloging}, we opted for a more innovative approach. Based on \texttt{NN\_FIRE}, a parallel model was constructed as shown in Figure~\ref{fig:NN_parallel}. The training methodology employed for this model was similar to that of the base model, but with a shorter training period. Furthermore, the \texttt{StandardScaler} was supplanted by a \texttt{RobustScaler} to facilitate improved normalization of the realistic data. All the weights of the base model are frozen during the training process and its output is concatenated with the output of a sub-network that takes $J_{R}$, $J_{z}$ and $J_{\phi}$ as input, which denotes the radial, vertical, and azimuthal actions, respectively (e.g. \citealt{binney2012actions, sanders2016review}). To enhance the normalization process, the logarithm of $J_{R}$ and $J_{z}$ was taken. The concatenated output will be further processed by the rest two dense layers, finally giving the classification result. The inputs of the sub-network introduce the potential of the MW into the parallel model and transform the model from a purely kinematic-based classifier to one that is based on dynamics. In this way, the parallel structure not only solves the extrapolation problem to a certain extent, but also maximizes the utilization of all available information of the six-dimensional parameters. We denote this model as \texttt{NN\_parallel} in the following text.

\subsubsection {Observational data}\label{section:sec2.2.1}
The third data release of $Gaia$ \citep{GaiaDR3} provided radial velocity measurement for over 30 million stars, thereby presenting an ideal target sample for our study. To overcome the limitation of parallax measurement, we employed the photo-astrometric distance estimated by \citet{anders2022photo}. We adopted stars with available radial velocity and photo-astrometric distance within $Gaia$ DR3, and imposed certain restrictions to ensure the quality of the data, which include a renormalized unit weight error (RUWE) $<$ 1.4 (e.g. \citealt{lindegren2021gaia}), a radial velocity error $<$ 20 km s$^{-1}$, and a distance error $<$ 30\%.
\par
In order to derive robust dynamic parameters, we employ a Monte Carlo methodology. For each observational parameter, we perform 100 times of random sampling, which follows a Gaussian distribution with the observed quantity serving as the mean and the error of the observed quantity as the standard deviation. In this way, each star is represented by 100 sets of Monte Carlo observations, which are then transformed into Cartesian positions and velocities. We apply a right-hand Galactocentric Cartesian coordinate system $(x, y, z)$, where the $x$-axis is oriented from the Sun towards the Galactic center, the $y$-axis aligns with the direction of the MW’s rotation, and the $z$-axis points towards the North Galactic Pole. We assume the circular speed at the Sun as 232.8 km s$^{-1}$ and the Solar Galactocentric distance as 8.2 kpc \citep{mcmillan2016mass}. The distance from the Sun to the Galactic plane is set at 20.8 pc \citep{Bennett19}. For the solar peculiar motion, we adopt the value reported by \citet{schonrich2010local}, yielding (U, V, W) $=$ (11.1, 12.24, 7.25) km s$^{-1}$. Subsequently, we calculate the corresponding actions and orbital parameters using \texttt{AGAMA} \citep{vasiliev2019agama}, with the assumption of \citet{mcmillan2016mass} potential and the application of \texttt{St\"ackel} approximation as described by \citet{binney2012actions}. The median values of the dynamic parameters are adopted as the final estimation.
\par
It is important to note that in the region behind the Galactic centre and close to the Galactic plane, the distances of a small number of stars may be overestimated. Since \citet{anders2022photo} utilized the extinction curve provided by \citet{schlafly2016optical}, whose fitting quality for the inner galaxy is relatively inferior (see Figure 10 in \citealt{schlafly2016optical}), we consider that this overestimation could potentially be attributed to extinction issues. Consequently, we remove stars in the region defined by: $x$ $>$ 0 kpc, $z$ $>$ -3 kpc, r$_{gc}$ $>$ 6 kpc, -50$^{\circ}$ $<$ $l$ $<$ 50$^{\circ}$ and -10$^{\circ}$ $<$ $b$ $<$ 15$^{\circ}$. Here, r$_{gc}$ denotes the Galactocentric distance, while $l$ and $b$ represent the Galactic longitude and Galactic latitude, respectively. Given the asymmetric distribution of the stars whose distances may be overestimated, we adopt an asymmetric criterion. Furthermore, we exclude stars with positive total energy, leaving a target sample of 27 085 748 stars, which is comparable in magnitude to the sample size of stars with 5D phase space parameters in \citet{ostdiek2020cataloging}.

\subsubsection {Chemical tagging of the training samples}\label{section:sec2.2.2}
For the purpose of training the \texttt{NN\_parallel} model to be applicable to the target sample, it is necessary to construct a new dataset based on observational data. The primary challenge in this process is determining an appropriate method for labeling the observational data. \citet{ostdiek2020cataloging} implemented a selection criterion, referred to as the ZM method (e.g. \citealt{herzog2018empirical}), where stars with [Fe/H] $< -1.5$ dex and $|z| > 1.5$ kpc were labeled as ex-situ. Nevertheless, given that the input of our NN consists solely of dynamic parameters, it is desirable to label ex-situ and in-situ stars using a criterion that is independent of dynamics, which ensures that the sample remains dynamically unbiased. Thus the measurement of chemical abundance is considered to be the ideal criterion for the tagging of ex-situ components in observations, as it offers an independent perspective, and serves as a reliable indicator of stellar origin.
\par
In pursuit of a large sample of stars with detailed chemical abundances, we cross-matched the $Gaia$ DR3 target sample with APOGEE DR17 catalogue \citep{accetta2022seventeenth}. Adhering to the online documentation of APOGEE DR17 and \citet{holtzman2015abundances}, we only adopted stars that satisfy the following criteria: 
\begin{enumerate}
\item No BAD bit in \texttt{STARFLAG} or \texttt{ASPCAPFLAG}, 
\item 3500 K $<$ $T\rm{_{eff}}$ $<$ 6000 K, 
\item S/N $\geq$ 70, 
\item VSCATTER $<$ 1 km $s^{-1}$, 
\item \texttt{X\_Fe\_FLAG} $==$ 0 and \texttt{X\_FE\_ERR} $<$ 0.15 dex. 
\end{enumerate}
Upon applying these selection parameters, a total of 254 046 stars remained. Additionally, we utilized the value-added catalogue (VAC) of LAMOST DR8 presented by \citet{li2022stellar}, which contains abundance information for 10 elements. The elemental abundances were estimated by an NN trained on APOGEE DR17 data and show high consistency with APOGEE measurements. Therefore, we further adopted 562 626 stars with \texttt{X\_FE\_ERR} $<$ 0.10 dex in the VAC that are in common with our target sample as a supplement to the dataset, resulting in a final sample size of 816 672 stars.
\par
As demonstrated in \citet{hawkins2015using}, the Galactic components can be labeled using [Al/Fe] and [Mg/Mn]. Magnesium is the first element to be affected by Type II supernovae, while manganese is produced in higher fractions compared to iron in Type Ia supernovae. Aluminium, on the other hand, is produced by Type II supernovae and is sensitive to the initial abundance of carbon and nitrogen, which originate from helium burning or Asymptotic Giant Branch evolution. As a result, the [Mg/Mn] and [Al/Fe] distributions of ex-situ stars are expected to differ from those of stars formed locally in the MW due to differences in their star formation rate and chemical evolution history. Based on APOGEE DR14 \citep{abolfathi2018fourteenth}, \citet{das2020ages} introduced the [Mg/Mn]-[Al/Fe] diagram and discovered a 'blob' of accreted stars located in the region of high [Mg/Mn] and low [Al/Fe], which is distinct from the stellar disc. \citet{horta2022chemical} and \citet{ortigoza2023Galactic} also characterized numerous substructures distributed within this region, including the GSE, Sequoia, and Heracles. Furthermore, recent studies have attempted to extract the ex-situ components in the [Mg/Mn]-[Al/Fe] diagram (e.g., \citealt{horta2021evidence, carrillo2023can, feltzing2023metal}). However, most of these efforts have focused primarily on the GSE region. As noted by \citet{hasselquist2021apogee} and \citet{fernandes2023comparative}, satellite galaxies of the MW such as the Large Magellanic Cloud (LMC), the Small Magellanic Cloud (SMC) and the Sgr exhibit relatively low [Mg/Mn] values. With the aim of uniformly distinguishing between the ex-situ and in-situ components, it may be more effective to select a region that encompasses as many dwarf galaxies and substructures as possible, rather than focusing on a specific accretion component such as the GSE.
\par
\begin{figure*}
 \includegraphics[width=\textwidth]{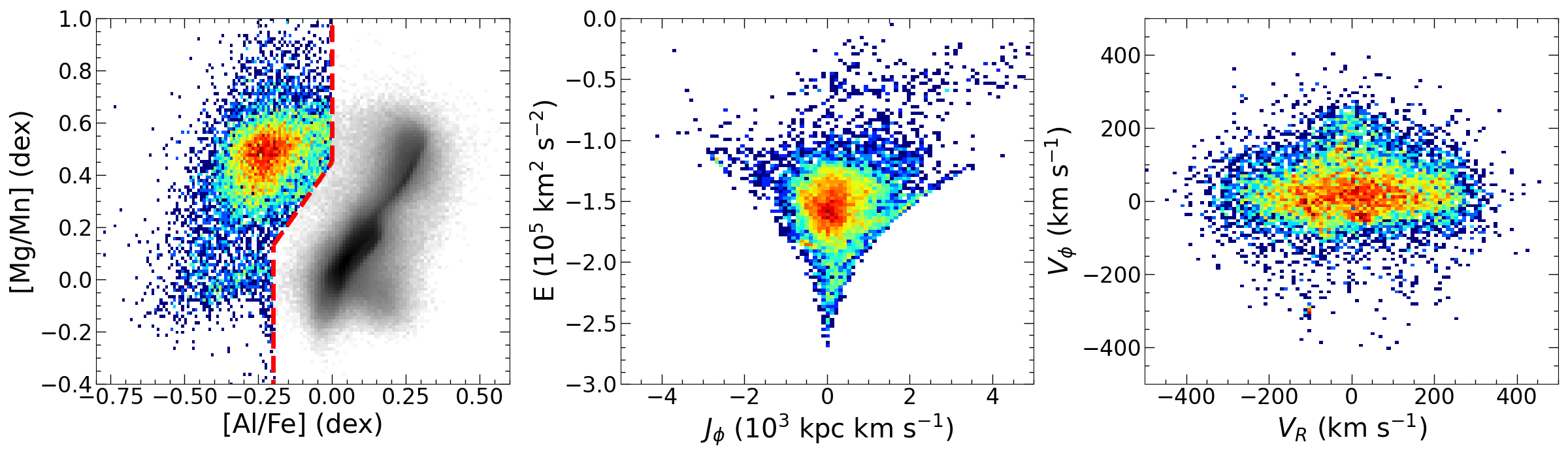}
 \caption{As depicted by the red dashed line in the [Mg/Mn]-[Al/Fe] plane, our criterion divides the plane into two distinct regions. The region to the left of the line is predominantly occupied by ex-situ stars, while the region to the right is primarily composed of in-situ stars. The middle and right subplots illustrate the distribution of stars labeled as ex-situ in $E-J_{\Phi}$ space and velocity space, respectively. Each subplot is colour-coded to represent the number density of stars, with warmer colours indicating higher densities for ex-situ stars and darker colours indicating higher densities for in-situ stars.}
 \label{fig:chemical_tagging}
\end{figure*}

We performed a segmental analysis of our training samples, as depicted in Figure~\ref{fig:chemical_tagging}. Stars were labeled as ex-situ if they satisfy the following selection criteria:
\begin{equation}\label{con:criteria}
\begin{cases}
[{\rm Al/Fe}]<0.00 & \text{if } [{\rm Mg/Mn}]\geq0.45 \\
[{\rm Mg/Mn}]>1.6[{\rm Al/Fe}]+0.45 & \text{if } 0.13\leq [{\rm Mg/Mn}]<0.45 \\
[{\rm Al/Fe}]<-0.20 & \text{if } [{\rm Mg/Mn}]<0.13
\end{cases}
\end{equation}
According to these criteria, 11 478 ex-situ stars were selected in our dataset. This selection includes member stars of the MW’s satellite galaxies and GC debris (e.g. \citealt{fernandez2022Galactic}). Inevitably, it also encompasses a minor fraction of stars associated with the Galactic disc. As shown in Figure~\ref{fig:chemical_tagging}, there is a group of stars with a $v_\Phi$ of approximately 230 km s$^{-1}$, whose distribution on the $E-J_{\Phi}$ plane closely resembles that of the disc. \citet{fernandes2023comparative} showed that these disc stars occupy the region of low [Al/Fe] and low [Mg/Mn], which overlaps with the ex-situ region as defined by Equation~\ref{con:criteria}. However, a criterion stringent enough to exclude these disc stars would also fail to capture the region where the majority of dwarf galaxies reside. Upon careful evaluation, we determined that this classification method is reasonable and the pollution of disc stars will not have a significant impact on the training of \texttt{NN\_parallel}.
\par

\subsubsection{Model performance}\label{section:sec2.2.3}
Since the weights of the base model are frozen, the chemical criterion turns out to be sufficient and robust enough to support the subsequent training of the \texttt{NN\_parallel}. During the training phase, the training samples were divided into training set, validation set and test set as in Section~\ref{section:sec2.1.2}. On the test set, \texttt{NN\_parallel} achieved an AUC of 0.98, which is comparable to that of the \texttt{NN\_FIRE}. Employing a default threshold value of 0.5, the NN attained a precision of 81.0\% and a recall of 64.6\%, which already signified an equilibrium between purity and completeness. This threshold will be utilized in subsequent analysis. Nevertheless, it is imperative to note that the metric values should be regarded as intuitive references only, as they are not predicated on truth-level labels as explicated in Section~\ref{section:sec2.1.2}.
\par
\begin{figure*}
 \includegraphics[width=\textwidth]{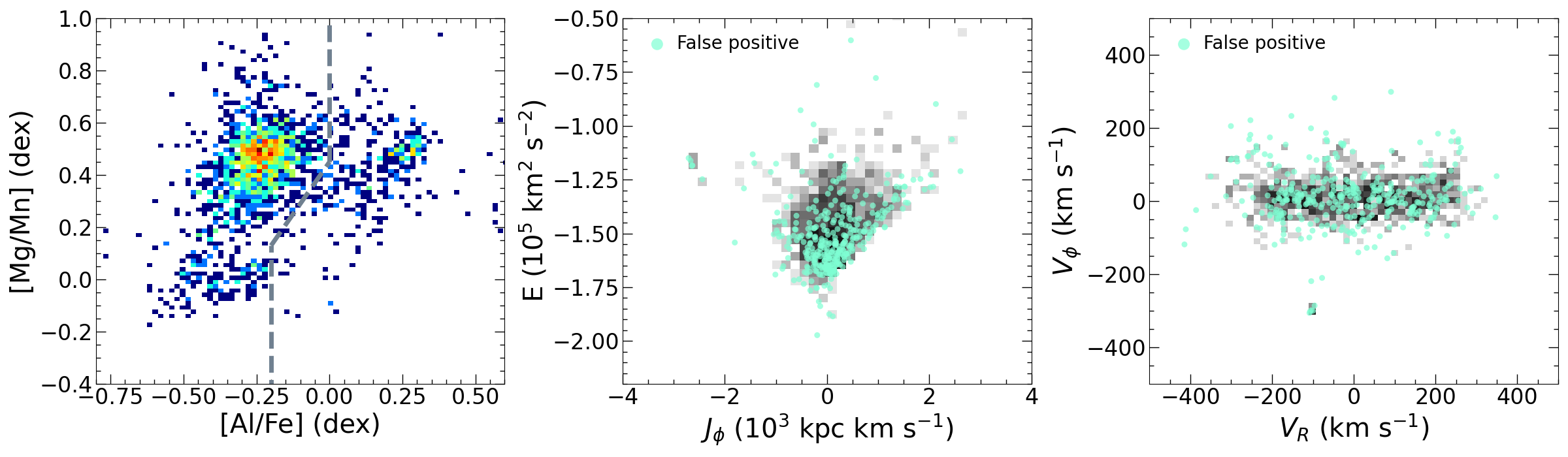}
 \caption{Ex-situ stars selected by the \texttt{NN\_parallel} model in the test set. The left panel displays the distribution in the [Mg/Mn]-[Al/Fe] diagram, with the selection criterion indicated by a grey dashed line. Samples located in the right part are considered false positive samples with respect to the chemical tagged label. The middle and right subplots show the distribution in $E-J_{\Phi}$ space and velocity space, respectively. The background of these two subplots represents the number density of ex-situ stars selected by the NN, with darker colours indicating higher densities. False positive samples are highlighted for emphasis, constituting 0.2\% of the samples in the test set.}
 \label{fig:test_set_result}
\end{figure*}

Independent of metrics, we conducted a more comprehensive analysis of the classification results given by \texttt{NN\_parallel}. As depicted in Figure~\ref{fig:test_set_result}, the selection made by the parallel model does not precisely mirror the pattern of the label. The \texttt{NN\_parallel} avoids selecting the disc stars that was incorrectly labeled as ex-situ through chemical tagging. Conversely, stars selected by the NN but labeled as in-situ through chemical tagging exhibit chemical patterns and dynamic characteristics consistent with those of the thick disc and the GSE.
Since the catalogue in \citet{sanderson2020synthetic} did not contain information about [Al/Fe] or [Mn/Fe], we were not able to reproduce the workflow of the second training phase on the simulation. However, based on the results depicted in Figure~\ref{fig:test_set_result}, it appears that \texttt{NN\_parallel} may offer an even more appropriate classification result than chemical tagging. A similar situation arose in \citet{ostdiek2020cataloging}, who randomly selected 200 000 synthetic stars from m12f and labeled them using the traditional ZM selection method. They then performed transfer learning on an NN pre-trained on m12i using this dataset. After transfer learning, the NN achieved a precision of 41\% and a recall of 47\% when the threshold was set at 0.75. When the threshold was raised to 0.95, the precision increased to 59\% while the recall decreased to 13\%. In comparison, the ZM selection method only achieved a precision of 50.9\% and a recall of 2.4\%. Our second training phase, like transfer learning, is designed to retain the results of the first training phase while adapting the model to observational data. The training process enabled the parallel model to overcome the limitations of both the base model and chemical tagging, resulting in improved performance. So far, the \texttt{NN\_parallel} model is well-prepared for the classification task of the target sample.

\section{Result}\label{section:sec3}
\begin{figure*}
 \includegraphics[width=\textwidth]{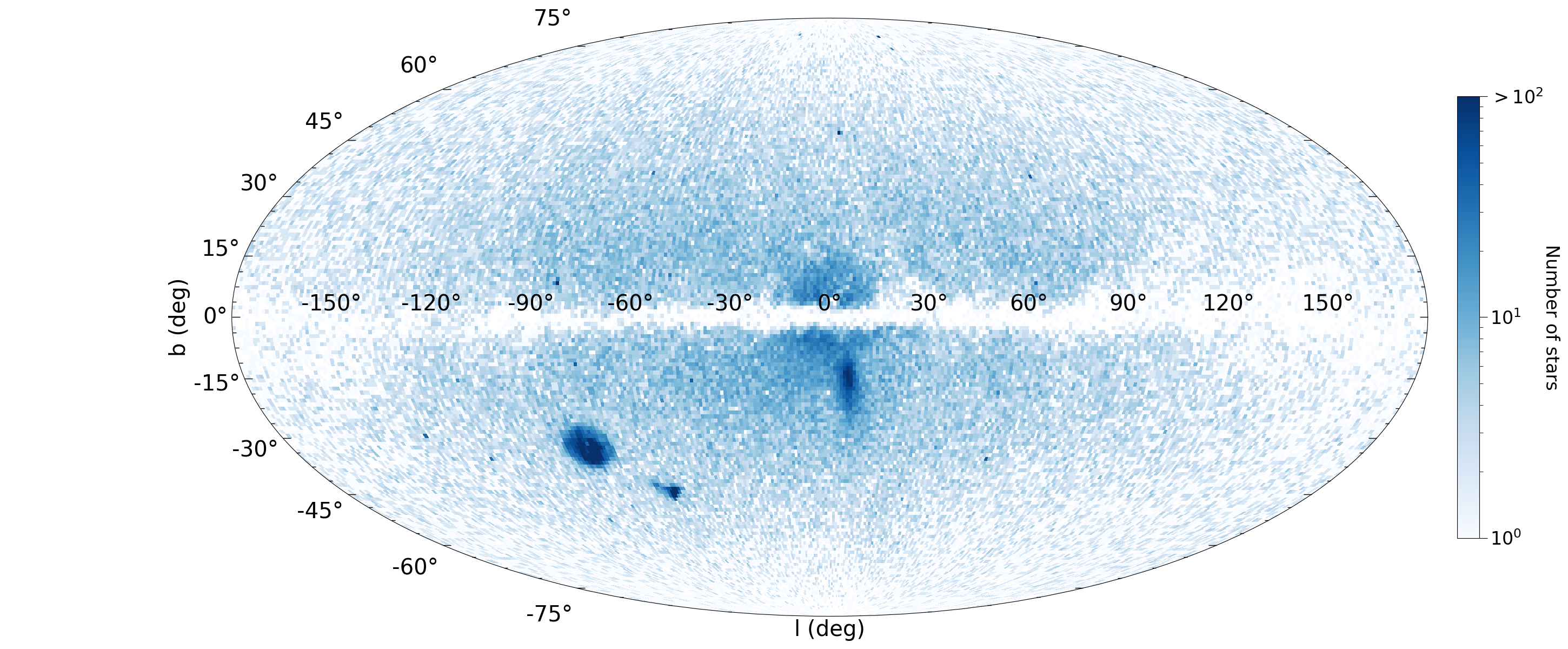}
 \caption{Spatial distribution of 160 146 ex-situ stars, as selected by the \texttt{NN\_parallel} model, is presented within the Galactic coordinate system. Three prominent overdensities are identified, corresponding to the LMC, SMC, and Sgr, respectively. Furthermore, globular clusters can be found as smaller overdensities. A total of 20 globular clusters are present in our ex-situ sample.}
 \label{fig:gl_gb}
\end{figure*}

After two phases of training, the \texttt{NN\_parallel} model was applied to the target sample from $Gaia$ DR3, resulting in the identification of 160 146 ex-situ stars. As illustrated in Figure~\ref{fig:gl_gb}, the region of the stellar disc was virtually eliminated, with the majority of the ex-situ stars being diffusely distributed in the rest of the plane. Specifically, three prominent spacial overdensities can be intuitively identified, corresponding to the LMC, SMC, and Sgr, respectively. Although not as conspicuous as dwarf galaxies, smaller overdensities in the plot reveal the presence of globular clusters, with a total of 20 being found. The further selection for dwarf galaxies and globular clusters is detailed in Section~\ref{section:sec3.1}. Substructures, which are considered to be disrupted dwarf galaxies that merged into the MW, cannot be directly identified in the coordinate space but are crucial for understanding the MW’s merger history. The selected member stars from substructures including the GSE, Thamnos, Sequoia, Helmi streams, Wukong, and Pontus are presented in Section~\ref{section:sec3.2}. As exhibited in Table~\ref{tab:Sample_size}, we summarize the sample size of the dwarf galaxies, globular clusters and substructures in the subsequent discussion.
\begin{table}
	\centering
	\caption{Sample size of the three dwarf galaxies, 20 globular clusters, and six substructures included in Section~\ref{section:sec3}.}
	\label{tab:Sample_size}
	\begin{tabular*}{\columnwidth}{@{\extracolsep{\fill}}lc}
		\hline
		Structures & N stars\\
		\hline
		LMC & 9822\\
  	    SMC & 1899\\
		Sgr core & 3319\\
		Sgr stream & 1269\\
		Globular Clusters & 831\\
		GSE & 47 521\\
		Thamnos & 5291\\
  	    Sequoia & 1714\\
		Helmi streams & 1655\\
		Wukong & 3140\\
		Pontus & 668\\
		\hline
            Total & 77 129\\
            \hline
	\end{tabular*}
\end{table}

\begin{figure*}
 \includegraphics[width=\textwidth]{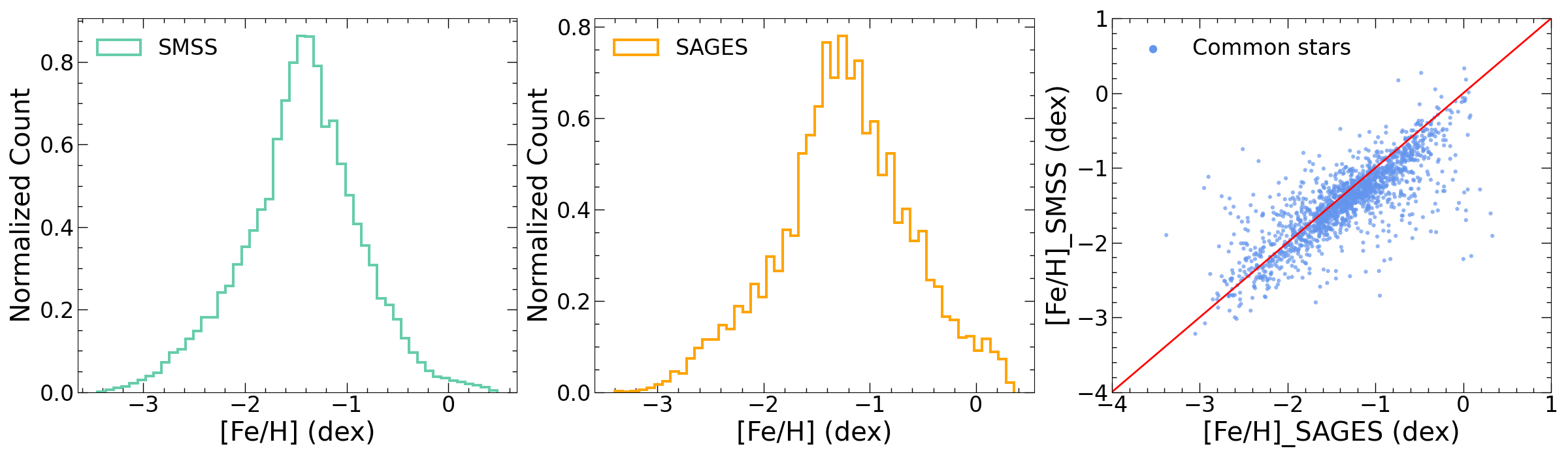}
 \caption{The left subplot presents the metallicity distribution of 62 025 ex-situ stars derived from SMSS, while the middle subplot exhibits the metallicity distribution of 18 413 ex-situ stars derived from SAGES. On the right, the subplot illustrates the correlation of iron abundance measurements for 1581 common stars between the SMSS and SAGES samples, with the red line indicating a one-to-one correlation. Despite slight system differences, the result demonstrate a good consistency.} 
 \label{fig:metallicity_distribution}
\end{figure*}

Furthermore, we examined the metallicity distribution of the ex-situ sample utilizing the metallicity derived from SkyMapper Southern Survey (SMSS; \citealt{wolf2018skymapper, onken2019skymapper, huang2022beyond}) and the Stellar Abundance and Galactic Evolution Survey (SAGES; \citealt{fan2023stellar, huang2023spectroscopy}). The SMSS is a photometric survey of the southern sky, while the SAGES covers the northern sky. By cross-matching our ex-situ sample with these two catalogues, we obtained 62 025 and 18 413 common stars, respectively. As shown in the left column of Figure~\ref{fig:metallicity_distribution}, the ex-situ stars exhibits a peak metal abundance around -1.4 dex. The peak metallicity of the SAGES sample is slightly higher at approximately -1.3 dex. We checked the agreement of metal abundance measurements for the two samples using 1581 common stars between them, resulting in a good consistency. However, the measurements of the SAGES samples were slightly higher than those of the SMSS samples, possibly due to differences in the filters utilized by the two surveys. Consequently, the peak metallicity of the ex-situ sample is likely to lie between -1.3 and -1.4 dex, which is consistent with the typical metallicity of GSE (e.g., \citealt{myeong2019evidence, Thetaleofthetail, GSEWWB}). Taking into account both metallicity analysis and star counts as shown in Table~\ref{tab:Sample_size}, our ex-situ sample appears to be composed of multiple components but predominated by the GSE.

\subsection{Dwarf galaxies and globular clusters}\label{section:sec3.1}
\begin{figure*}
 \includegraphics[width=\textwidth]{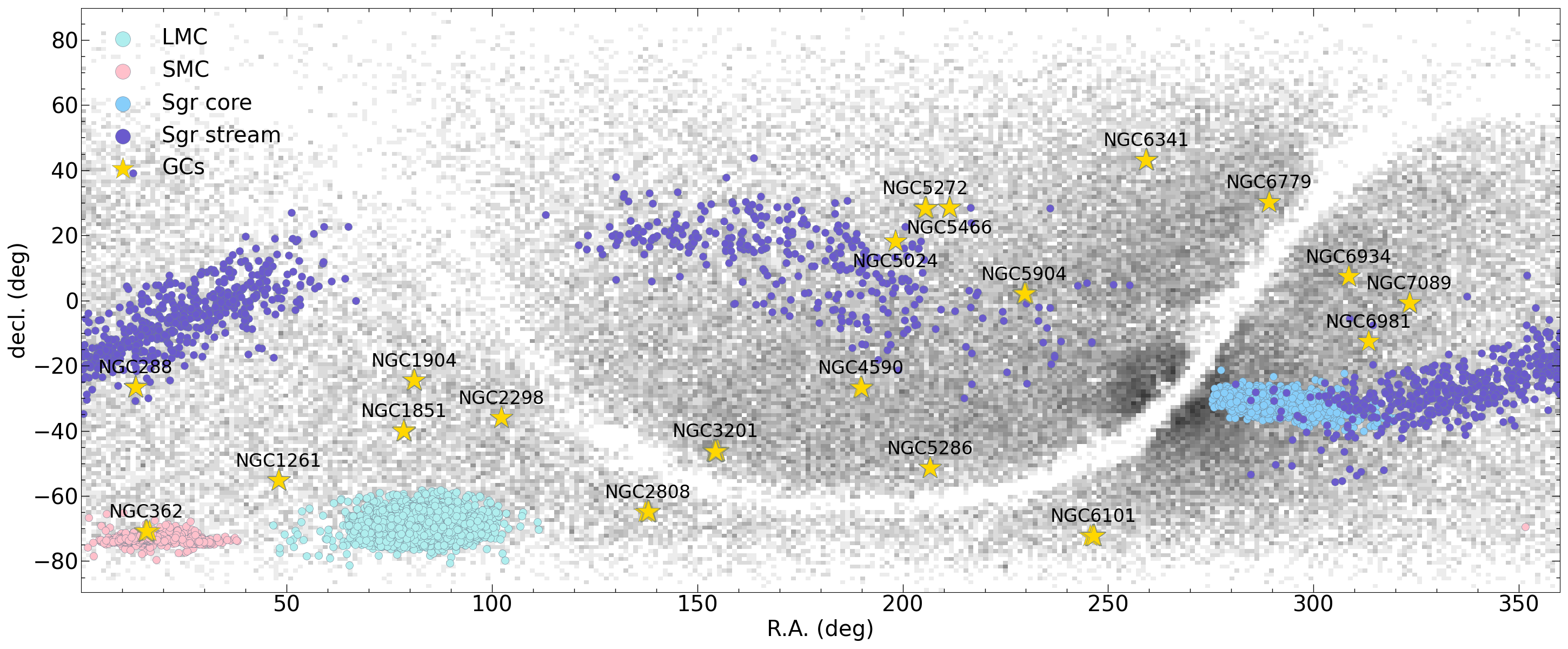}
 \caption{Distribution of the LMC, SMC, Sgr core, Sgr stream and globular clusters on the celestial sphere. Dwarf galaxies are depicted in distinct colours, and the names of globular clusters are annotated near their corresponding positions. The background illustrates the number density of the ex-situ sample.}
 \label{fig:ra_dec}
\end{figure*}

In our ex-situ sample, dwarf galaxies (DGs) and globular clusters (GCs) are the most prominent structures, exhibiting a distinct clustering in coordinate space. As shown in Table~\ref{tab:DGs}, we selected three DGs combining the position and velocity. Moreover, we selected the disrupted portion of the Sgr by cross-matching with Sgr stream catalogues compiled by \citet{YCQ2019} and \citet{Sgrstream}. They are shown here as a comparison of the Sgr core. The member stars of these four systems are plotted on the celestial sphere in Figure~\ref{fig:ra_dec}, where the LMC, SMC, and Sgr core exhibit distinct spatial clustering, while the Sgr stream displays a more extended distribution. Thereafter, we removed the selected member stars of the DGs from our sample, which made the overdensity regions of the GCs more prominent, facilitating the subsequent selection.
\par
For each globular cluster, we primarily select its member stars using the centre coordinate and angular size ($\theta$) from references listed in Table~\ref{tab:GCs}. We retain stars within a projected distance ($d_{\rm{proj}}$) to the globular cluster centre of $d_{\rm{proj}}<1.5\times \theta$, where $1.5\times \theta$ corresponds to three times the angular radii. Subsequently, we verify whether the selected globular cluster candidates exhibit similar motion properties by examining their proper motion and radial velocity. Outliers in proper motion phase space are manually eliminated, and stars with radial velocity outside the range of $\mu_{\rm{rv}}\pm3\sigma_{\rm{rv}}$ are further removed. Here, $\mu_{\rm{rv}}$ represents the median radial velocity of a globular cluster, and $\sigma_{\rm{rv}}$ is half the difference between the corresponding value of the 84\textsuperscript{th} percentile and 16\textsuperscript{th} percentile. We only selected GCs with more than 10 member stars. This resulted in a total of 20 GCs, all of which are classified as ex-situ in \citet{belokurov2023situ}.
\par
The metric values, such as precision and recall, provide an indication of the average classification ability of the overall sample. While the NN demonstrates a high degree of accuracy in identifying certain portions of the sample, its performance may be somewhat diminished when classifying more challenging portions. As dwarf galaxies and globular clusters exhibit distinct spatial clustering, their identification by the \texttt{NN\_parallel} represents a relatively minor test of its capabilities. The true potential of the NN is further demonstrated by its ability to accurately identify member stars of substructures that no longer cluster in coordinate space.

\subsection{Substructures}\label{section:sec3.2}
Prior to the selection of substructures, we excluded the member stars of DGs and GCs as identified in Section~\ref{section:sec3.1}, which was necessary to prevent any overlap with the regions where substructures reside in the phase space. During the selection process, we adopted an approach similar to that of \citet{naidu2020evidence}, where each time member stars of a substructure was selected, we removed them from the sample to prevent their influence on the subsequent selection of other substructures. In total, 143 006 ex-situ stars were included in our analysis. We then selected 59 989 member stars from six substructures according to the order and criteria outlined in Table~\ref{tab:Substructures}. 
\par
\begin{figure*}
 \includegraphics[width=\textwidth]{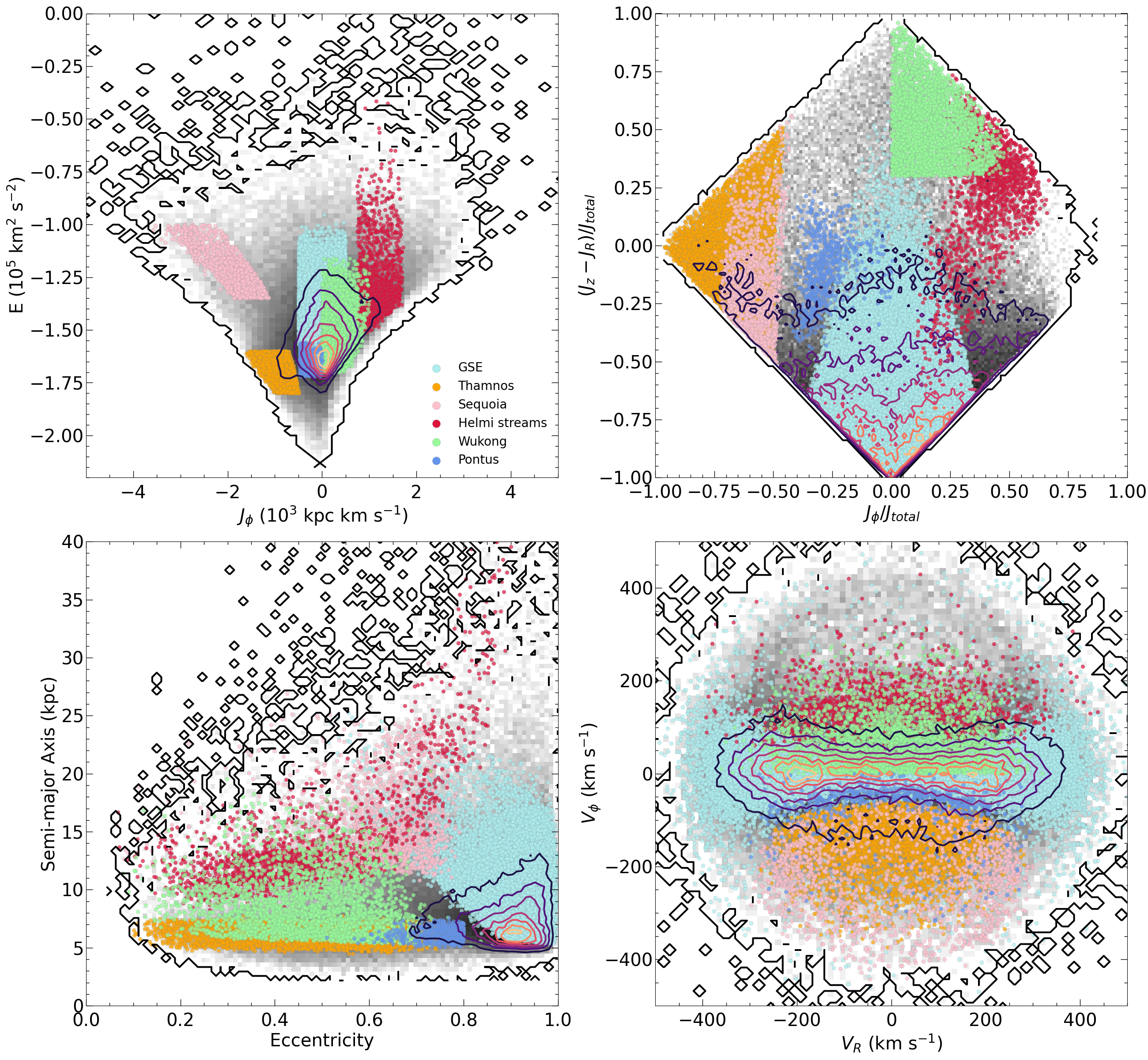}
 \caption{The member stars of the six substructures, as selected using the criteria outlined in Table~\ref{tab:Substructures}, are depicted in four distinct phase spaces. Each substructure is represented by a unique colour. The background and the contours illustrate the number density of the remaining 143 006 ex-situ stars after the removal of dwarf galaxies and globular clusters.}
 \label{fig:substructures}
\end{figure*}

Our selection focuses on substructures including the GSE, Thamnos, Sequoia, Helmi streams, Wukong (also known as LMS-1), and Pontus. Figure~\ref{fig:substructures} displays the distribution of their member stars in four phase spaces. The second subplot presents an action diamond, with $J\rm_{total}$ defined as the sum of $J_{R}$, $J_{z}$, and $|J_{\Phi}|$ (e.g., \citealt{myeong2019evidence}). In the third subplot, we characterize the semi-major axis as $\rm(r_{apo}+r_{peri})/2$, and the eccentricity as $\rm(r_{apo}-r_{peri})/(r_{apo}+r_{peri})$, where $\rm r_{apo}$ and $\rm r_{peri}$ represent the distances to the apocentre and pericentre, respectively. The selection criteria, as listed in Table~\ref{tab:Substructures}, involve multi-dimensional linear cuts. Figure~\ref{fig:substructures} illustrates the projection of samples with high-dimensional features in four two-dimensional spaces. As a result, the substructures may display clear linear boundaries in the corresponding plane.

\subsection{Ex-situ percentage}\label{section:sec3.3}
\begin{figure}
 \includegraphics[width=\columnwidth]{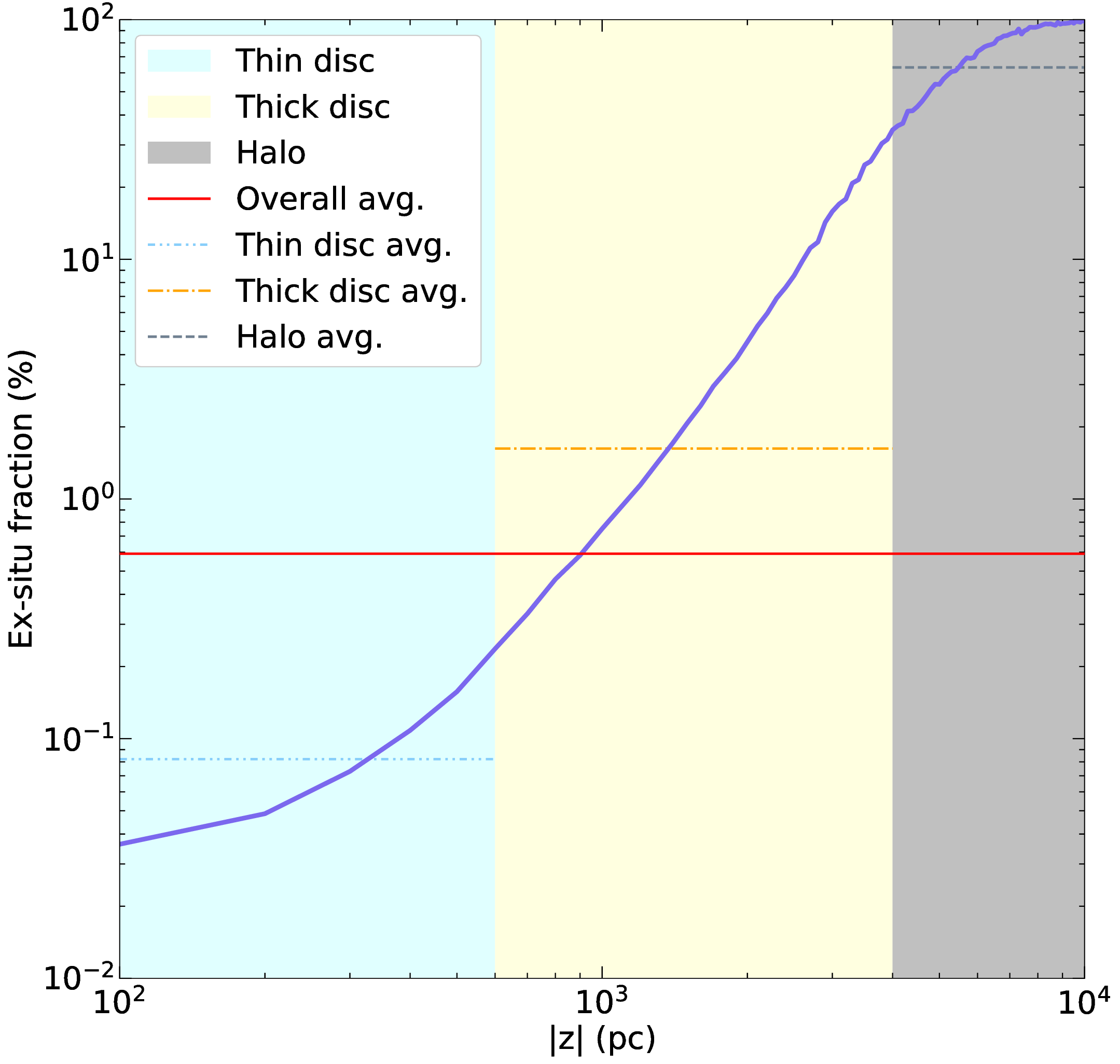}
 \caption{Ex-situ fraction as a function of $\lvert z \rvert$. The region with $\lvert z \rvert$ below 0.6 kpc is considered as the thin disc, while the region with $\lvert z \rvert$ between 0.6 kpc and 4 kpc is considered as the thick disc. The region above 4 kpc and below 10 kpc is occupied by the halo. Different regions are represented in distinct colours, and the average ex-situ fraction is indicated by horizontal dashed lines in the corresponding region. The red horizontal lines indicate the average ex-situ fraction of the target sample.}
 \label{fig:exsitu_percentage}
\end{figure}

In the presence of a large sample of high purity ex-situ stars, we are able to conduct a statistical analysis of the ex-situ fraction in the MW. To ensure the robustness of our analysis, we focus exclusively on target samples located in the region where $|z|\leq$ 10 kpc. In Figure~\ref{fig:exsitu_percentage_2D}, we present a two-dimensional ex-situ fraction map that exhibits significant layering along the $|z|$-direction. Based on the layering pattern, we divide our sample into three distinct regions, corresponding to the thin disc ($|z| < 0.6$ kpc), the thick disc ($0.6$ kpc $\leq |z| < 4$ kpc), and the halo (4 kpc $\leq |z| \leq 10$ kpc). The ex-situ samples, as identified by the \texttt{NN\_parallel} model, were binned according to a bin size of $|z| =$ 100 pc. We then plotted the ex-situ fraction as a function of $|z|$ in Figure~\ref{fig:exsitu_percentage}, where the $x$-axis represents the right boundary of each bin and the $y$-axis stands for the corresponding ex-situ fraction. Figure~\ref{fig:exsitu_percentage} provides a clear illustration of how the proportion of ex-situ stars varies with the vertical height from the Galactic disc. It also shows the proportion of ex-situ stars in the three components of the MW. The average ex-situ percentages of the thin disc, the thick disc, and the halo are 0.1\%, 1.6\%, and 63.2\% \footnote{The estimations for both the thin disc and the thick disc yield standard errors less than 0.01\%, while for the halo, the standard error is 0.02\%. All estimated values are rounded to one decimal places for consistency.}, respectively. Consistent with our previous knowledge, the thin disc is predominantly composed of in-situ stars. In the region of the thick disc, there is a substantial increase in the proportion of ex-situ stars, while on average, the thick disc is still primarily composed of in-situ stars. In contrast, the stellar halo is dominated by ex-situ stars, with this dominance increasing with $|z|$.
\par
\begin{figure}
 \includegraphics[width=\columnwidth]{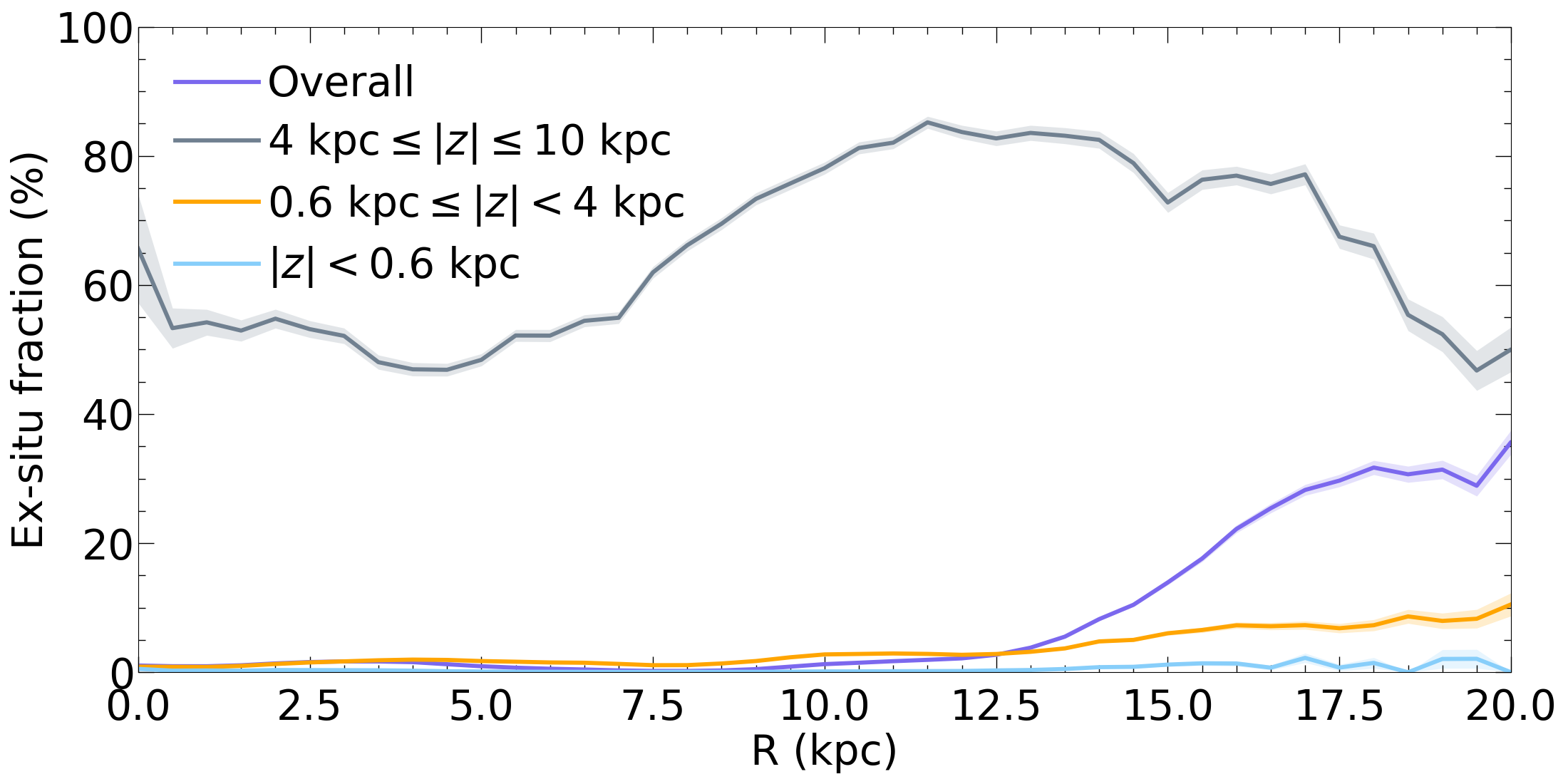}
 \caption{Ex-situ fraction as a function of R is illustrated by the purple line, which shows the overall trend of the sample. Distributions for the thin disc, thick disc, and halo as defined in Figure~\ref{fig:exsitu_percentage} are depicted in distinct colours. The standard error of each bin is indicated by the shaded area.}
 \label{fig:exsitu_percentage_R}
\end{figure}

Figure~\ref{fig:exsitu_percentage_R} presents an alternative perspective in the Galactocentric cylindrical coordinate system, illustrating the variation in the proportion of ex-situ stars with respect to their distance from the Galactic centre (R). The data is binned using a bin size of R $=$ 500 pc, and the $x$-axis represents the left boundary of each bin. As illustrated by the purple line in Figure~\ref{fig:exsitu_percentage_R}, the fraction of ex-situ stars is less than 2\% up until R $\approx 12$ kpc, and it significantly increases towards the outskirts of the MW. A general growth in the ex-situ fraction with increasing R can also be observed in both the thick disc and thin disc, with the trend being more pronounced in the thick disc. The ex-situ fraction in the thick disc is higher than that in the thin disc over the entire coverage of R, which is also indicated by the average values in Figure~\ref{fig:exsitu_percentage}. Within the halo, ex-situ stars hold a dominant position and the curve is mainly influenced by the spatial distribution of member stars from distinct substructures. When R is less than 5 kpc, the ex-situ fraction is relatively stable. As R extends beyond 5 kpc, the trend of the curve begins to be dominated by the GSE (e.g \citealt{an2021blueprint, GSEWWB}), exhibiting an upward trajectory until R approximates 12 kpc. Subsequently, the curve commences its decline, aligning with the spatial distribution trend of GSE member stars (e.g. \citealt{naidu2021reconstructing}).

\section{Discussion}\label{section:sec4}
\subsection{Threshold setting}
In Section~\ref{section:sec3}, all the ex-situ stars are selected using a threshold value of 0.5. This threshold was chosen as it maintains an optimal balance between precision and recall, as evidenced by the results in the test set. However, as the metrics are not based on truth-level label, our choice can only be taken as a reference. In practice, the setting of threshold may vary depending on the scientific goals. If the goal is to obtain a purer ex-situ sample, the threshold can be increased. On the other hand, if a purer in-situ sample is desired, stars with small raw prediction values from the NN should be selected. Additionally, the adjustment to threshold value can reduce the impact of samples with inaccurate observations to a certain extent. By combining a relatively stringent threshold with a restriction on the error of observations, the quality of the required samples can be effectively ensured. 
\par
If we take one step further, it becomes apparent that the ultimate goal of both traditional linear segmentation methods (e.g., \citealt{nissen2010two}) and deep learning methods is to obtain an ex-situ sample that conforms to experience and previous knowledge. The primary difference between the two methods lies in their performance. As demonstrated in \citet{ostdiek2020cataloging}, traditional methods struggle to balance precision and recall, with high recall often accompanied by low precision. Although deep learning methods are not currently capable of breaking this deadlock, they are able to achieve higher recall under the same precision level compared with traditional methods, and vice versa. Under certain threshold values, an NN model can even surpass traditional methods in both metrics. The dynamic classification of stars involves many dimensions of information, and NN excels at processing high-dimensional data. Traditional selection methods directly restrict dynamic parameters, providing clear physical meaning. However, this approach may overlook certain dimensions and fail to account for potential connections between parameters of different dimensions. The NN method, on the other hand, directly outputs a continuous value between 0 and 1. All we need to do is determine a threshold to divide these continuous values into two categories. While adjusting the threshold may appear less physically meaningful than adjusting limits on dynamic parameters, it is important to note that the physical meaning is already reflected in the mapping established by the NN from high-dimensional dynamic parameters to output values. Adjusting the threshold is simply an intuitive and efficient process to obtain the final classification result. Moreover, the threshold is a hyperparameter that is independent of the model structure, enabling the attainment of different classification outcomes with out necessitating any modifications to the NN itself. By fine-tuning the threshold, classification results can be tailored to achieve various scientific objectives and meet specific requirements.

\subsection{Result assessment}
In contrast to simulations, it is extremely difficult to perfectly distinguish between in-situ and ex-situ stars in observations. Consequently, validating the ex-situ sample identified by \texttt{NN\_parallel} presents a significant challenge, and we have refrained from quantitatively evaluating the performance of the NN. Nevertheless, as depicted in Figure~\ref{fig:gl_gb}, the NN has effectively eliminated the majority of the disc stars. This is further evidenced by Figure~\ref{fig:substructures}, which clearly shows that the area where the disc is located has been accurately deducted. In Section~\ref{section:sec3}, we present a selection of member stars from dwarf galaxies, globular clusters, and substructures within our ex-situ sample. The member stars are selected based on the criteria established in the literature, without the use of any clustering algorithm. This approach was chosen as a quick and convenient means to demonstrate the potential of our algorithm and to verify the reliability of our ex-situ samples. It should be noted that most of the selection methods employed are relatively conservative. As such, the selected member stars are not exhaustive, as evidenced by the remaining overdensities in Figure~\ref{fig:substructures}. In forthcoming studies, clustering algorithms will be employed to facilitate more comprehensive investigations and detailed examination of each substructure. Furthermore, the origin of the substructures and potential correlations between various substructures will be investigated through simulations (e.g. \citealt{amarante2022gastro}). 
\par
In Section~\ref{section:sec3.3}, we provided a comprehensive illustration of the variation in the proportion of ex-situ stars with respect to the vertical height from the Galactic disc and the distance from the Galactic centre. However, it is important to note that our results are subject to the selection effect of $Gaia$ and the performance of the NN. As such, the result we present only reflect the analysis of our current sample. We divide our sample into three distinct parts according to the layering pattern in Figure~\ref{fig:exsitu_percentage_2D}, which do not precisely align with the three components of the MW as defined canonically \citep{bland2016galaxy}. The estimation of the proportion of ex-situ stars in the thin disc, thick disc, and halo should be considered as preliminary. What we present is far from a definitive conclusion but rather a possibility for statistical exploration of the ex-situ components within the MW. Despite the current limitations, we believe that our study provides novel insights and serves as a catalyst for future research endeavors.

\section{Conclusion}\label{section:sec5}
In this work, we developed a deep learning methodology to identify the ex-situ stars within $Gaia$ DR3. Our base model, referred to as \texttt{NN\_FIRE}, was built to learn fundamental kinematic features. The training of \texttt{NN\_FIRE} was conducted on a synthetic $Gaia$ catalogue derived from a Milky Way-mass galaxy from the $Latte$ suite of FIRE-2 simulations, specifically the m12i. The input for \texttt{NN\_FIRE} consists of 3D position and velocity data, which are transformed from \texttt{ra}, \texttt{dec}, \texttt{pmra}, \texttt{pmdec}, \texttt{radial\_velocity}, and \texttt{dhel\_true}. The output of the NN is a series of continuous values ranging from 0 to 1, which can be interpreted as the probability of a star being ex-situ. According to the truth-level label, the \texttt{NN\_FIRE} model achieved an AUC exceeding 0.98 on the test set. At a threshold of 0.5, we obtained a precision of 80.1\% and a recall of 34.5\%. By increasing the threshold to 0.75, the precision escalated to as high as 98.3\%. 
\par
Based on \texttt{NN\_FIRE}, a senior model noted as \texttt{NN\_parallel} was built, which processes dynamic data derived from real observations of $Gaia$. During the training phase of \texttt{NN\_parallel}, the weights of the base model were frozen. The output of the base model was concatenated with the output of a sub-network that processes $J_{R}$, $J_{z}$ and $J_{\Phi}$. We adopted stars with available radial velocity and photo-astrometric distance estimated by \citet{anders2022photo} within $Gaia$ DR3 as our target sample and cross-matched it with the APOGEE DR17 catalogue and LAMOST DR8 VAC to build a dataset with elemental abundance information for training \texttt{NN\_parallel}. The stars were labeled using a chemical tagging method, as outlined in Section~\ref{section:sec2.2.2}, which performed a segmental selection in the [Mg/Mn]-[Al/Fe] plane. Although the chemical tagging method did mix some disc stars into the ex-situ sample, with the aid of the first training phase, \texttt{NN\_parallel} overcame this problem and produced better results than selecting solely through chemical criteria.
\par
Out of the 27 085 748 stars in the target sample, the \texttt{NN\_parallel} model successfully identified 160 146 ex-situ stars. According to the metallicity distribution, the ex-situ sample is composed of multiple components, with the GSE being the most prominent. A significant number of member stars from the LMC, SMC, Sgr, as well as 20 globular clusters were found in the ex-situ sample. Additionally, member stars from substructures including the GSE, Thamnos, Sequoia, Helmi streams, Wukong, and Pontus were also selected. The identification of these member stars not only verifies the reliability of our ex-situ sample and our algorithm but also provides a substantial sample for future research on dwarf galaxies, globular clusters, and substructures. Moreover, we provided an intuitive illustration demonstrating an increase in the ex-situ fraction as the vertical height from the Galactic disc and the distance from the Galactic centre increase. Finally, We conducted a preliminary estimation of the proportion of ex-situ stars in the thin disc, thick disc, and halo, yielding percentages of 0.1\%, 1.6\%, and 63.2\%, respectively.
\par
This study serves as a prior work, with the primary objective of introducing our proposed algorithm and providing a reliable large sample of ex-situ stars. The findings delineated herein are the product of a preliminary analysis of our ex-situ sample. While these findings do not encompass the full scope of potential results, we posit that they effectively underscore the viability and promise of our proposed methodology. It is our aspiration that this research will bring the community fresh insights and inspiration regarding the evolution and merger history of the MW. Looking ahead, we plan to augment our research by incorporating simulations, conducting chemical abundance analyses, and employing clustering algorithms to delve deeper into our findings. We eagerly anticipate unearthing new revelations in this exciting field of study.

\section*{Acknowledgements}
This study is supported by the National Natural Science Foundation of China under grant Nos. 11988101, 11890694, 12373020, National Key R\&D Program of China Nos. 2019YFA0405500, 2023YFE0107800 and CAS Project for Young Scientists in Basic Research grant No. YSBR-062. ZL thanks Drs. Haining Li, and Wenbo Wu for their helpful discussions and insightful suggestions. JA acknowledges funding from the European Research Council (ERC) under the European Union’s Horizon 2020 research and innovation programme (grant agreement No. 852839). We have made use of data from the European Space Agency (ESA) mission
{\it Gaia} (\url{https://www.cosmos.esa.int/gaia}), processed by the {\it Gaia}
Data Processing and Analysis Consortium (DPAC,
\url{https://www.cosmos.esa.int/web/gaia/dpac/consortium}). Funding for the DPAC
has been provided by national institutions, in particular the institutions
participating in the {\it Gaia} Multilateral Agreement.
\par
Guoshoujing Telescope (the Large Sky Area Multi-Object Fiber Spectroscopic Telescope LAMOST) is a National Major Scientific Project built by the Chinese Academy of Sciences. Funding for the project has been provided by the National Development and Reform Commission. LAMOST is operated and managed by the National Astronomical Observatories, Chinese Academy of Sciences.
\par
Funding for the Sloan Digital Sky 
Survey IV has been provided by the 
Alfred P. Sloan Foundation, the U.S. 
Department of Energy Office of 
Science, and the Participating 
Institutions.
SDSS-IV acknowledges support and 
resources from the Center for High 
Performance Computing  at the 
University of Utah. The SDSS 
website is \url{www.sdss.org}.

SDSS-IV is managed by the 
Astrophysical Research Consortium 
for the Participating Institutions 
of the SDSS Collaboration including 
the Brazilian Participation Group, 
the Carnegie Institution for Science, 
Carnegie Mellon University, Center for 
Astrophysics | Harvard \& 
Smithsonian, the Chilean Participation 
Group, the French Participation Group, 
Instituto de Astrof\'isica de 
Canarias, The Johns Hopkins 
University, Kavli Institute for the 
Physics and Mathematics of the 
Universe (IPMU) / University of 
Tokyo, the Korean Participation Group, 
Lawrence Berkeley National Laboratory, 
Leibniz Institut f\"ur Astrophysik 
Potsdam (AIP),  Max-Planck-Institut 
f\"ur Astronomie (MPIA Heidelberg), 
Max-Planck-Institut f\"ur 
Astrophysik (MPA Garching), 
Max-Planck-Institut f\"ur 
Extraterrestrische Physik (MPE), 
National Astronomical Observatories of 
China, New Mexico State University, 
New York University, University of 
Notre Dame, Observat\'ario 
Nacional / MCTI, The Ohio State 
University, Pennsylvania State 
University, Shanghai 
Astronomical Observatory, United 
Kingdom Participation Group, 
Universidad Nacional Aut\'onoma 
de M\'exico, University of Arizona, 
University of Colorado Boulder, 
University of Oxford, University of 
Portsmouth, University of Utah, 
University of Virginia, University 
of Washington, University of 
Wisconsin, Vanderbilt University, 
and Yale University. \par

\section*{Data Availability}
The catalogue of NN classification scores is available at Zenodo: \url{https://doi.org/10.5281/zenodo.20378639}. The simulated data utilized during the training phase is accessible at \url{http://ananke.hub.yt}. Astrometric and chemical abundance data can be obtained from the websites of $Gaia$ (\url{https://gea.esac.esa.int/archive/}), APOGEE (\url{https://www.sdss4.org/dr17/irspec/dr_synopsis}) and LAMOST (\url{http://www.lamost.org/dr8/v1.1/doc/vac}), respectively. The result of this study will be shared upon reasonable request to the corresponding author.


\bibliographystyle{mnras}
\bibliography{References}

@article{steinmetz2006radial,
  title={The radial velocity experiment (RAVE): first data release},
  author={Steinmetz, Matthias and Zwitter, Toma{\v{z}} and Siebert, Arnaud and Watson, Fred G and Freeman, Kenneth C and Munari, Ulisse and Campbell, Rachel and Williams, Mary and Seabroke, George M and Wyse, Rosemary FG and others},
  journal={The Astronomical Journal},
  volume={132},
  number={4},
  pages={1645},
  year={2006},
  publisher={IOP Publishing}
}

@article{yanny2009segue,
  title={SEGUE: a spectroscopic survey of 240,000 stars with g= 14--20},
  author={Yanny, Brian and Rockosi, Constance and Newberg, Heidi Jo and Knapp, Gillian R and Adelman-McCarthy, Jennifer K and Alcorn, Bonnie and Allam, Sahar and Allende, Carlos Prieto and An, Deokkeun and Anderson, Kurt SJ and others},
  journal={The Astronomical Journal},
  volume={137},
  number={5},
  pages={4377},
  year={2009},
  publisher={IOP Publishing}
}

@article{deng2012lamost,
  title={LAMOST Experiment for Galactic Understanding and Exploration (LEGUE)—The survey's science plan},
  author={Deng, Li-Cai and Newberg, Heidi Jo and Liu, Chao and Carlin, Jeffrey L and Beers, Timothy C and Chen, Li and Chen, Yu-Qin and Christlieb, Norbert and Grillmair, Carl J and Guhathakurta, Puragra and others},
  journal={Research in Astronomy and Astrophysics},
  volume={12},
  number={7},
  pages={735},
  year={2012},
  publisher={IOP Publishing}
}

@article{liu2013lss,
  title={LSS-GAC--A LAMOST Spectroscopic Survey of the Galactic Anti-center},
  author={Liu, X-W and Yuan, H-B and Huo, Z-Y and Deng, L-C and Hou, J-L and Zhao, Y-H and Zhao, G and Shi, J-R and Luo, A-L and Xiang, M-S and others},
  journal={Proceedings of the International Astronomical Union},
  volume={9},
  number={S298},
  pages={310--321},
  year={2013},
  publisher={Cambridge University Press}
}

@article{zhao2012lamost,
  title={LAMOST spectral survey—An overview},
  author={Zhao, Gang and Zhao, Yong-Heng and Chu, Yao-Quan and Jing, Yi-Peng and Deng, Li-Cai},
  journal={Research in Astronomy and Astrophysics},
  volume={12},
  number={7},
  pages={723},
  year={2012},
  publisher={IOP Publishing}
}

@article{de2015galah,
  title={The GALAH survey: scientific motivation},
  author={De Silva, Gayandhi M and Freeman, Ken C and Bland-Hawthorn, Jonathan and Martell, Sarah and De Boer, E Wylie and Asplund, Martin and Keller, Stefan and Sharma, Sanjib and Zucker, Daniel B and Zwitter, Tomaz and others},
  journal={Monthly Notices of the Royal Astronomical Society},
  volume={449},
  number={3},
  pages={2604--2617},
  year={2015},
  publisher={Oxford University Press}
}

@article{majewski2017apache,
  title={The apache point observatory galactic evolution experiment (APOGEE)},
  author={Majewski, Steven R and Schiavon, Ricardo P and Frinchaboy, Peter M and Prieto, Carlos Allende and Barkhouser, Robert and Bizyaev, Dmitry and Blank, Basil and Brunner, Sophia and Burton, Adam and Carrera, Ricardo and others},
  journal={The Astronomical Journal},
  volume={154},
  number={3},
  pages={94},
  year={2017},
  publisher={IOP Publishing}
}

@article{lecun2015deep,
  title={Deep learning},
  author={LeCun, Yann and Bengio, Yoshua and Hinton, Geoffrey},
  journal={nature},
  volume={521},
  number={7553},
  pages={436--444},
  year={2015},
  publisher={Nature Publishing Group}
}

@article{abolfathi2018fourteenth,
  title={The fourteenth data release of the Sloan Digital Sky Survey: First spectroscopic data from the extended Baryon Oscillation Spectroscopic Survey and from the second phase of the Apache Point Observatory Galactic Evolution Experiment},
  author={Abolfathi, Bela and Aguado, DS and Aguilar, Gabriela and Prieto, Carlos Allende and Almeida, Andres and Ananna, Tonima Tasnim and Anders, Friedrich and Anderson, Scott F and Andrews, Brett H and Anguiano, Borja and others},
  journal={The Astrophysical Journal Supplement Series},
  volume={235},
  number={2},
  pages={42},
  year={2018},
  publisher={IOP Publishing}
}

@article{luo2015first,
  title={The first data release (DR1) of the LAMOST regular survey},
  author={Luo, A-Li and Zhao, Yong-Heng and Zhao, Gang and Deng, Li-Cai and Liu, Xiao-Wei and Jing, Yi-Peng and Wang, Gang and Zhang, Hao-Tong and Shi, Jian-Rong and Cui, Xiang-Qun and others},
  journal={Research in Astronomy and Astrophysics},
  volume={15},
  number={8},
  pages={1095},
  year={2015},
  publisher={IOP Publishing}
}

@article{accetta2022seventeenth,
  title={The Seventeenth Data Release of the Sloan Digital Sky Surveys: Complete Release of MaNGA, MaStar, and APOGEE-2 Data},
  author={Accetta, Katherine and Aerts, Conny and Aguirre, Victor Silva and Ahumada, Romina and Ajgaonkar, Nikhil and Ak, N Filiz and Alam, Shadab and Prieto, Carlos Allende and Almeida, Andr{\'e}s and Anders, Friedrich and others},
  journal={The Astrophysical Journal Supplement Series},
  volume={259},
  number={2},
  pages={35},
  year={2022},
  publisher={IOP Publishing}
}

@article{das2020ages,
  title={Ages and kinematics of chemically selected, accreted Milky Way halo stars},
  author={Das, Payel and Hawkins, Keith and Jofr{\'e}, Paula},
  journal={Monthly Notices of the Royal Astronomical Society},
  volume={493},
  number={4},
  pages={5195--5207},
  year={2020},
  publisher={Oxford University Press}
}

@article{buder2021galah+,
  title={The GALAH+ survey: Third data release},
  author={Buder, Sven and Sharma, Sanjib and Kos, Janez and Amarsi, Anish M and Nordlander, Thomas and Lind, Karin and Martell, Sarah L and Asplund, Martin and Bland-Hawthorn, Joss and Casey, Andrew R and others},
  journal={Monthly Notices of the Royal Astronomical Society},
  volume={506},
  number={1},
  pages={150--201},
  year={2021},
  publisher={Oxford University Press}
}

@article{zhao2006stellar,
  title={Stellar abundance and Galactic chemical evolution through LAMOST spectroscopic survey},
  author={Zhao, Gang and Chen, Yu-Qin and Shi, Jian-Rong and Liang, Yan-Chun and Hou, Jin-Liang and Chen, Li and Zhang, Hua-Wei and Li, Ai-Gen},
  journal={Chinese Journal of Astronomy and Astrophysics},
  volume={6},
  number={3},
  pages={265},
  year={2006},
  publisher={IOP Publishing}
}

@article{li2015spectroscopic,
  title={Spectroscopic analysis of metal-poor stars from LAMOST: early results},
  author={Li, Hai-Ning and Zhao, Gang and Christlieb, Norbert and Wang, Liang and Wang, Wei and Zhang, Yong and Hou, Yonghui and Yuan, Hailong},
  journal={The Astrophysical Journal},
  volume={798},
  number={2},
  pages={110},
  year={2015},
  publisher={IOP Publishing}
}

@article{zhao2009catalog,
  title={A Catalog of Moving Group Candidates in The Solar Neighborhood},
  author={Zhao, Jingkun and Zhao, Gang and Chen, Yuqin},
  journal={The Astrophysical Journal},
  volume={692},
  number={2},
  pages={L113},
  year={2009},
  publisher={IOP Publishing}
}

@article{helmi1999debris,
  title={Debris streams in the solar neighbourhood as relicts from the formation of the Milky Way},
  author={Helmi, Amina and White, Simon DM and De Zeeuw, P Tim and Zhao, HongSheng},
  journal={Nature},
  volume={402},
  number={6757},
  pages={53--55},
  year={1999},
  publisher={Nature Publishing Group}
}

@article{myeong2019evidence,
  title={Evidence for two early accretion events that built the Milky Way stellar halo},
  author={Myeong, GC and Vasiliev, E and Iorio, G and Evans, NW and Belokurov, V},
  journal={Monthly Notices of the Royal Astronomical Society},
  volume={488},
  number={1},
  pages={1235--1247},
  year={2019},
  publisher={Oxford University Press}
}

@article{helmi2020streams,
  title={Streams, substructures, and the early history of the Milky Way},
  author={Helmi, Amina},
  journal={Annual Review of Astronomy and Astrophysics},
  volume={58},
  pages={205--256},
  year={2020},
  publisher={Annual Reviews}
}

@article{helmi2018merger,
  title={The merger that led to the formation of the Milky Way’s inner stellar halo and thick disk},
  author={Helmi, Amina and Babusiaux, Carine and Koppelman, Helmer H and Massari, Davide and Veljanoski, Jovan and Brown, Anthony GA},
  journal={Nature},
  volume={563},
  number={7729},
  pages={85--88},
  year={2018},
  publisher={Nature Publishing Group}
}

@article{belokurov2018co,
  title={Co-formation of the disc and the stellar halo},
  author={Belokurov, V and Erkal, Denis and Evans, NW and Koposov, SE and Deason, AJ},
  journal={Monthly Notices of the Royal Astronomical Society},
  volume={478},
  number={1},
  pages={611--619},
  year={2018},
  publisher={Oxford University Press}
}

@article{horta2022chemical,
  title={The chemical characterisation of halo substructure in the Milky Way based on APOGEE},
  author={Horta, Danny and Schiavon, Ricardo P and Mackereth, J Ted and Weinberg, David H and Hasselquist, Sten and Feuillet, Diane and O’Connell, Robert W and Anguiano, Borja and Allende-Prieto, Carlos and Beaton, Rachael L and others},
  journal={Monthly Notices of the Royal Astronomical Society},
  year={2022}
}

@article{ostdiek2020cataloging,
  title={Cataloging accreted stars within Gaia DR2 using deep learning},
  author={Ostdiek, Bryan and Necib, Lina and Cohen, Timothy and Freytsis, Marat and Lisanti, Mariangela and Garrison-Kimmmel, S and Wetzel, Andrew and Sanderson, Robyn E and Hopkins, Philip F},
  journal={Astronomy \& Astrophysics},
  volume={636},
  pages={A75},
  year={2020},
  publisher={EDP Sciences}
}

@article{li2022stellar,
  title={The stellar parameters and elemental abundances from low-resolution spectra--I. 1.2 million giants from LAMOST DR8},
  author={Li, Zhuohan and Zhao, Gang and Chen, Yuqin and Liang, Xilong and Zhao, Jingkun},
  journal={Monthly Notices of the Royal Astronomical Society},
  volume={517},
  number={4},
  pages={4875--4891},
  year={2022},
  publisher={Oxford University Press}
}

@article{mcmillan2016mass,
  title={The mass distribution and gravitational potential of the Milky Way},
  author={McMillan, Paul J},
  journal={Monthly Notices of the Royal Astronomical Society},
  volume={465},
  pages={76--94},
  year={2017},
  publisher={Oxford University Press}
}

@article{binney2012actions,
  title={Actions for axisymmetric potentials},
  author={Binney, James},
  journal={Monthly Notices of the Royal Astronomical Society},
  volume={426},
  number={2},
  pages={1324--1327},
  year={2012},
  publisher={Blackwell Science Ltd Oxford, UK}
}

@article{schonrich2010local,
  title={Local kinematics and the local standard of rest},
  author={Sch{\"o}nrich, Ralph and Binney, James and Dehnen, Walter},
  journal={Monthly Notices of the Royal Astronomical Society},
  volume={403},
  number={4},
  pages={1829--1833},
  year={2010},
  publisher={The Royal Astronomical Society}
}

@article{white1978core,
  title={Core condensation in heavy halos: a two-stage theory for galaxy formation and clustering},
  author={White, Simon DM and Rees, Martin J},
  journal={Monthly Notices of the Royal Astronomical Society},
  volume={183},
  number={3},
  pages={341--358},
  year={1978},
  publisher={Oxford University Press Oxford, UK}
}

@article{ibata1994dwarf,
  title={A dwarf satellite galaxy in Sagittarius},
  author={Ibata, Rodrigo A and Gilmore, Gerry and Irwin, MJ},
  journal={Nature},
  volume={370},
  number={6486},
  pages={194--196},
  year={1994},
  publisher={Nature Publishing Group UK London}
}

@article{myeong2018sausage,
  title={The sausage globular clusters},
  author={Myeong, GC and Evans, NW and Belokurov, V and Sanders, JL and Koposov, SE},
  journal={The Astrophysical Journal Letters},
  volume={863},
  number={2},
  pages={L28},
  year={2018},
  publisher={IOP Publishing}
}

@article{gaia2018gaia,
  title={Gaia Data Release 2 Summary of the contents and survey properties},
  author={{Gaia Collaboration} and Brown, AGA and Vallenari, A and Prusti, T and De Bruijne, JHJ and Babusiaux, C and Juh{\'a}sz, {\'A}L and Marschalk{\'o}, G and Marton, G and Moln{\'a}r, L and others},
  journal={Astronomy \& Astrophysics},
  volume={616},
  number={1},
  year={2018},
  publisher={EDP Sciences}
}

@article{haywood2018disguise,
  title={In disguise or out of reach: first clues about in situ and accreted stars in the stellar halo of the Milky Way from Gaia DR2},
  author={Haywood, Misha and Di Matteo, Paola and Lehnert, MD and Snaith, Owain and Khoperskov, Sergey and G{\'o}mez, Ana},
  journal={The Astrophysical Journal},
  volume={863},
  number={2},
  pages={113},
  year={2018},
  publisher={IOP Publishing}
}

@article{gallart2019uncovering,
  title={Uncovering the birth of the Milky Way through accurate stellar ages with Gaia},
  author={Gallart, Carme and Bernard, Edouard J and Brook, Chris B and Ruiz-Lara, Tom{\'a}s and Cassisi, Santi and Hill, Vanessa and Monelli, Matteo},
  journal={Nature Astronomy},
  volume={3},
  number={10},
  pages={932--939},
  year={2019},
  publisher={Nature Publishing Group UK London}
}

@article{conroy2019mapping,
  title={Mapping the Stellar Halo with the H3 Spectroscopic Survey},
  author={Conroy, Charlie and Bonaca, Ana and Cargile, Phillip and Johnson, Benjamin D and Caldwell, Nelson and Naidu, Rohan P and Zaritsky, Dennis and Fabricant, Daniel and Moran, Sean and Rhee, Jaehyon and others},
  journal={The Astrophysical Journal},
  volume={883},
  number={1},
  pages={107},
  year={2019},
  publisher={IOP Publishing}
}

@article{naidu2020evidence,
  title={Evidence from the h3 survey that the stellar halo is entirely comprised of substructure},
  author={Naidu, Rohan P and Conroy, Charlie and Bonaca, Ana and Johnson, Benjamin D and Ting, Yuan-Sen and Caldwell, Nelson and Zaritsky, Dennis and Cargile, Phillip A},
  journal={The Astrophysical Journal},
  volume={901},
  number={1},
  pages={48},
  year={2020},
  publisher={IOP Publishing}
}

@article{horta2021evidence,
  title={Evidence from APOGEE for the presence of a major building block of the halo buried in the inner Galaxy},
  author={Horta, Danny and Schiavon, Ricardo P and Mackereth, J Ted and Pfeffer, Joel and Mason, Andrew C and Kisku, Shobhit and Fragkoudi, Francesca and Allende Prieto, Carlos and Cunha, Katia and Hasselquist, Sten and others},
  journal={Monthly Notices of the Royal Astronomical Society},
  volume={500},
  number={1},
  pages={1385--1403},
  year={2021},
  publisher={Oxford University Press}
}

@article{yan2022overview,
  title={Overview of the LAMOST survey in the first decade},
  author={Yan, Hongliang and Li, Haining and Wang, Song and Zong, Weikai and Yuan, Haibo and Xiang, Maosheng and Huang, Yang and Xie, Jiwei and Dong, Subo and Yuan, Hailong and others},
  journal={The Innovation},
  pages={100224},
  year={2022},
  publisher={Elsevier}
}

@article{shih2022via,
  title={via machinae: Searching for stellar streams using unsupervised machine learning},
  author={Shih, David and Buckley, Matthew R and Necib, Lina and Tamanas, John},
  journal={Monthly Notices of the Royal Astronomical Society},
  volume={509},
  number={4},
  pages={5992--6007},
  year={2022},
  publisher={Oxford University Press}
}

@article{necib2020chasing,
  title={Chasing accreted structures within Gaia DR2 using deep learning},
  author={Necib, Lina and Ostdiek, Bryan and Lisanti, Mariangela and Cohen, Timothy and Freytsis, Marat and Garrison-Kimmel, Shea},
  journal={The Astrophysical Journal},
  volume={903},
  number={1},
  pages={25},
  year={2020},
  publisher={IOP Publishing}
}

@article{sanderson2020synthetic,
  title={Synthetic gaia surveys from the fire cosmological simulations of milky way-mass galaxies},
  author={Sanderson, Robyn E and Wetzel, Andrew and Loebman, Sarah and Sharma, Sanjib and Hopkins, Philip F and Garrison-Kimmel, Shea and Faucher-Gigu{\`e}re, Claude-Andr{\'e} and Kere{\v{s}}, Du{\v{s}}an and Quataert, Eliot},
  journal={The Astrophysical Journal Supplement Series},
  volume={246},
  number={1},
  pages={6},
  year={2020},
  publisher={IOP Publishing}
}

@article{carrillo2023can,
  title={Can we really pick and choose? Benchmarking various selections of Gaia Enceladus/Sausage stars in observations with simulations},
  author={Carrillo, Andreia and Deason, Alis J and Fattahi, Azadeh and Callingham, Thomas M and Grand, Robert JJ},
  journal={arXiv preprint arXiv:2306.00770},
  year={2023}
}

@inproceedings{campello2013density,
  title={Density-based clustering based on hierarchical density estimates},
  author={Campello, Ricardo JGB and Moulavi, Davoud and Sander, J{\"o}rg},
  booktitle={Advances in Knowledge Discovery and Data Mining: 17th Pacific-Asia Conference, PAKDD 2013, Gold Coast, Australia, April 14-17, 2013, Proceedings, Part II 17},
  pages={160--172},
  year={2013},
  organization={Springer}
}

@article{fix1951discriminatory,
  title={Discriminatory Analysis, Nonparametric Discrimination: Consistency Properties},
  author={Fix, E. and Hodges, J. L.},
  journal={USAF School of Aviation Medicine},
  year={1951}
}

@article{dempster1977maximum,
  title={Maximum likelihood from incomplete data via the EM algorithm},
  author={Dempster, Arthur P and Laird, Nan M and Rubin, Donald B},
  journal={Journal of the royal statistical society: series B (methodological)},
  volume={39},
  number={1},
  pages={1--22},
  year={1977},
  publisher={Wiley Online Library}
}

@article{yuan2020dynamical,
  title={Dynamical relics of the ancient galactic halo},
  author={Yuan, Zhen and Myeong, GC and Beers, Timothy C and Evans, N Wyn and Lee, Young Sun and Banerjee, Projjwal and Gudin, Dmitrii and Hattori, Kohei and Li, Haining and Matsuno, Tadafumi and others},
  journal={The Astrophysical Journal},
  volume={891},
  number={1},
  pages={39},
  year={2020},
  publisher={IOP Publishing}
}

@article{yuan2018stargo,
  title={StarGO: A New Method to Identify the Galactic Origins of Halo Stars},
  author={Yuan, Zhen and Chang, Jiang and Banerjee, Projjwal and Han, Jiaxin and Kang, Xi and Smith, MC},
  journal={The Astrophysical Journal},
  volume={863},
  number={1},
  pages={26},
  year={2018},
  publisher={IOP Publishing}
}

@article{borsato2020identifying,
  title={Identifying stellar streams in Gaia DR2 with data mining techniques},
  author={Borsato, Nicholas W and Martell, Sarah L and Simpson, Jeffrey D},
  journal={Monthly Notices of the Royal Astronomical Society},
  volume={492},
  number={1},
  pages={1370--1384},
  year={2020},
  publisher={Oxford University Press}
}

@article{ester1996dbscan,
  title={A density-based algorithm for discovering clusters in large spatial databases with noise},
  author={Ester, Martin and Kriegel, Hans-Peter and Sander, Jörg and Xu, Xiaowei},
  journal={Kdd},
  volume={96},
  number={34},
  pages={226--231},
  year={1996}
}

@book{kohonen2001self,
  title={Self-organizing maps},
  author={Kohonen, Teuvo},
  volume={30},
  year={2001},
  publisher={Springer},
  address={Berlin},
  edition={3},
  series={Springer Series in Information Sciences}
}

@article{nachman2020anomaly,
  title={Anomaly detection with density estimation},
  author={Nachman, Benjamin and Shih, David},
  journal={Physical Review D},
  volume={101},
  number={7},
  pages={075042},
  year={2020},
  publisher={APS}
}

@article{grillmair2006detection,
  title={Detection of a 60°-long dwarf galaxy debris stream},
  author={Grillmair, Carl J},
  journal={The Astrophysical Journal},
  volume={645},
  number={1},
  pages={L37},
  year={2006},
  publisher={IOP Publishing}
}

@article{necib2020evidence,
  title={Evidence for a vast prograde stellar stream in the solar vicinity},
  author={Necib, Lina and Ostdiek, Bryan and Lisanti, Mariangela and Cohen, Timothy and Freytsis, Marat and Garrison-Kimmel, Shea and Hopkins, Philip F and Wetzel, Andrew and Sanderson, Robyn},
  journal={Nature Astronomy},
  volume={4},
  number={11},
  pages={1078--1083},
  year={2020},
  publisher={Nature Publishing Group UK London}
}

@article{price2018astropy,
  title={The astropy project: building an open-science project and status of the v2. 0 core package},
  author={Price-Whelan, Adrian M and Sip{\H{o}}cz, BM and G{\"u}nther, HM and Lim, PL and Crawford, SM and Conseil, S and Shupe, DL and Craig, MW and Dencheva, N and Ginsburg, A and others},
  journal={The Astronomical Journal},
  volume={156},
  number={3},
  pages={123},
  year={2018},
  publisher={IOP Publishing}
}

@misc{chollet2015keras,
  title={Keras},
  author={Chollet, Francois and others},
  year={2015},
  howpublished={\url{https://github.com/fchollet/keras}},
}

@misc{tensorflow2015-whitepaper,
  title={ {TensorFlow}: Large-Scale Machine Learning on Heterogeneous Systems},
  url={https://www.tensorflow.org/},
  note={Software available from tensorflow.org},
  author={Abadi, Mart\'{i}n and others},
  year={2015},
}

@article{hahnloser2000digital,
  title={Digital selection and analogue amplification coexist in a cortex-inspired silicon circuit},
  author={Hahnloser, Richard HR and Sarpeshkar, Rahul and Mahowald, Misha A and Douglas, Rodney J and Seung, H Sebastian},
  journal={Nature},
  volume={405},
  number={6789},
  pages={947--951},
  year={2000},
  publisher={Nature Publishing Group}
}

@article{fan2023stellar,
  title={The Stellar Abundances and Galactic Evolution Survey (SAGES)----I. General Description and the First Data Release (DR1)},
  author={Fan, Zhou and Zhao, Gang and Wang, Wei and Zheng, Jie and Zhao, Jingkun and Li, Chun and Chen, Yuqin and Yuan, Haibo and Li, Haining and Tan, Kefeng and others},
  journal={arXiv preprint arXiv:2306.15611},
  year={2023}
}

@article{wolf2018skymapper,
  title={SkyMapper southern survey: first data release (DR1)},
  author={Wolf, Christian and Onken, Christopher A and Luvaul, Lance C and Schmidt, Brian P and Bessell, Michael S and Chang, Seo-Won and Da Costa, Gary S and Mackey, Dougal and Martin-Jones, Tony and Murphy, Simon J and others},
  journal={Publications of the Astronomical Society of Australia},
  volume={35},
  pages={e010},
  year={2018},
  publisher={Cambridge University Press}
}

@article{onken2019skymapper,
  title={SkyMapper Southern Survey: second data release (DR2)},
  author={Onken, Christopher A and Wolf, Christian and Bessell, Michael S and Chang, Seo-Won and Da Costa, Gary S and Luvaul, Lance C and Mackey, Dougal and Schmidt, Brian P and Shao, Li},
  journal={Publications of the Astronomical Society of Australia},
  volume={36},
  pages={e033},
  year={2019},
  publisher={Cambridge University Press}
}

@article{huang2022beyond,
  title={Beyond spectroscopy. I. Metallicities, distances, and age estimates for over 20 million stars from SMSS DR2 and Gaia EDR3},
  author={Huang, Yang and Beers, Timothy C and Wolf, Christian and Lee, Young Sun and Onken, Christopher A and Yuan, Haibo and Shank, Derek and Zhang, Huawei and Wang, Chun and Shi, Jianrong and others},
  journal={The Astrophysical Journal},
  volume={925},
  number={2},
  pages={164},
  year={2022},
  publisher={IOP Publishing}
}

@misc{huang2023spectroscopy,
      title={Beyond spectroscopy. II. Stellar parameters for over twenty million stars in the northern sky from SAGES DR1 and Gaia DR3}, 
      author={Yang Huang and Timothy C. Beers and Hai-Bo Yuan and Ke-Feng Tan and Wei Wang and Jie Zheng and Chun Li and Young Sun Lee and Hai-Ning Li and Jing-Kun Zhao and Xiang-Xiang Xue and Yu-Juan Liu and Hua-Wei Zhang and Xue-Ang Sun and Ji Li and Hong-Rui Gu and Christian Wolf and Christopher A. Onken and Ji-Feng Liu and Zhou Fan and Gang Zhao},
      year={2023},
      eprint={2307.04469},
      archivePrefix={arXiv},
      primaryClass={astro-ph.GA}
}

@inproceedings{lin2017focal,
  title={Focal loss for dense object detection},
  author={Lin, Tsung-Yi and Goyal, Priya and Girshick, Ross and He, Kaiming and Doll{\'a}r, Piotr},
  booktitle={Proceedings of the IEEE international conference on computer vision},
  pages={2980--2988},
  year={2017}
}

@article{hopkins2018fire,
  title={FIRE-2 simulations: physics versus numerics in galaxy formation},
  author={Hopkins, Philip F and Wetzel, Andrew and Kere{\v{s}}, Du{\v{s}}an and Faucher-Gigu{\`e}re, Claude-Andr{\'e} and Quataert, Eliot and Boylan-Kolchin, Michael and Murray, Norman and Hayward, Christopher C and Garrison-Kimmel, Shea and Hummels, Cameron and others},
  journal={Monthly Notices of the Royal Astronomical Society},
  volume={480},
  number={1},
  pages={800--863},
  year={2018},
  publisher={Oxford University Press}
}

@article{wetzel2023public,
  title={Public data release of the FIRE-2 cosmological zoom-in simulations of galaxy formation},
  author={Wetzel, Andrew and Hayward, Christopher C and Sanderson, Robyn E and Ma, Xiangcheng and Angl{\'e}s-Alc{\'a}zar, Daniel and Feldmann, Robert and Chan, TK and El-Badry, Kareem and Wheeler, Coral and Garrison-Kimmel, Shea and others},
  journal={The Astrophysical Journal Supplement Series},
  volume={265},
  number={2},
  pages={44},
  year={2023},
  publisher={IOP Publishing}
}

@article{bellardini20223d,
  title={3D elemental abundances of stars at formation across the histories of Milky Way-mass galaxies in the FIRE simulations},
  author={Bellardini, Matthew A and Wetzel, Andrew and Loebman, Sarah R and Bailin, Jeremy},
  journal={Monthly Notices of the Royal Astronomical Society},
  volume={514},
  number={3},
  pages={4270--4289},
  year={2022},
  publisher={Oxford University Press}
}

@article{sharma2011galaxia,
  title={Galaxia: A code to generate a synthetic survey of the Milky Way},
  author={Sharma, Sanjib and Bland-Hawthorn, Joss and Johnston, Kathryn V and Binney, James},
  journal={The Astrophysical Journal},
  volume={730},
  number={1},
  pages={3},
  year={2011},
  publisher={IOP Publishing}
}

@article{pedregosa2011scikit,
  title={Scikit-learn: Machine learning in Python},
  author={Pedregosa, Fabian and Varoquaux, Ga{\"e}l and Gramfort, Alexandre and Michel, Vincent and Thirion, Bertrand and Grisel, Olivier and Blondel, Mathieu and Prettenhofer, Peter and Weiss, Ron and Dubourg, Vincent and others},
  journal={the Journal of machine Learning research},
  volume={12},
  pages={2825--2830},
  year={2011},
  publisher={JMLR. org}
}

@article{xue2011quantifying,
  title={Quantifying Kinematic Substructure in the Milky Way's Stellar Halo},
  author={Xue, Xiang-Xiang and Rix, Hans-Walter and Yanny, Brian and Beers, Timothy C and Bell, Eric F and Zhao, Gang and Bullock, James S and Johnston, Kathryn V and Morrison, Heather and Rockosi, Constance and others},
  journal={The Astrophysical Journal},
  volume={738},
  number={1},
  pages={79},
  year={2011},
  publisher={IOP Publishing}
}

@article{kingma2014adam,
  title={Adam: A method for stochastic optimization},
  author={Kingma, Diederik P and Ba, Jimmy},
  journal={arXiv preprint arXiv:1412.6980},
  year={2014}
}

@article{vasiliev2019agama,
  title={AGAMA: action-based galaxy modelling architecture},
  author={Vasiliev, Eugene},
  journal={Monthly Notices of the Royal Astronomical Society},
  volume={482},
  number={2},
  pages={1525--1544},
  year={2019},
  publisher={Oxford University Press}
}

@article{hawkins2015using,
  title={Using chemical tagging to redefine the interface of the Galactic disc and halo},
  author={Hawkins, Keith and Jofre, Paula and Masseron, Thomas and Gilmore, Gerry},
  journal={Monthly Notices of the Royal Astronomical Society},
  volume={453},
  number={1},
  pages={758--774},
  year={2015},
  publisher={Oxford University Press}
}

@article{fernandes2023comparative,
  title={A comparative analysis of the chemical compositions of Gaia-Enceladus/Sausage and Milky Way satellites using APOGEE},
  author={Fernandes, Laura and Mason, Andrew C and Horta, Danny and Schiavon, Ricardo P and Hayes, Christian and Hasselquist, Sten and Feuillet, Diane and Beaton, Rachael L and J{\"o}nsson, Henrik and Kisku, Shobhit and others},
  journal={Monthly Notices of the Royal Astronomical Society},
  volume={519},
  number={3},
  pages={3611--3622},
  year={2023},
  publisher={Oxford University Press}
}

@article{feltzing2023metal,
  title={The metal-weak Milky Way stellar disk hidden in the Gaia-Sausage-Enceladus debris: the APOGEE DR17 view},
  author={Feltzing, Sofia and Feuillet, Diane},
  journal={arXiv preprint arXiv:2303.00016},
  year={2023}
}

@article{herzog2018empirical,
  title={Empirical determination of dark matter velocities using metal-poor stars},
  author={Herzog-Arbeitman, Jonah and Lisanti, Mariangela and Madau, Piero and Necib, Lina},
  journal={Physical review letters},
  volume={120},
  number={4},
  pages={041102},
  year={2018},
  publisher={APS}
}

@ARTICLE{Nidever2020,
       author = {{Nidever}, David L. and {Hasselquist}, Sten and {Hayes}, Christian R. and {Hawkins}, Keith and {Povick}, Joshua and {Majewski}, Steven R. and {Smith}, Verne V. and {Anguiano}, Borja and {Stringfellow}, Guy S. and {Sobeck}, Jennifer S. and {Cunha}, Katia and {Beers}, Timothy C. and {Bestenlehner}, Joachim M. and {Cohen}, Roger E. and {Garcia-Hernandez}, D.~A. and {J{\"o}nsson}, Henrik and {Nitschelm}, Christian and {Shetrone}, Matthew and {Lacerna}, Ivan and {Allende Prieto}, Carlos and {Beaton}, Rachael L. and {Dell'Agli}, Flavia and {Fern{\'a}ndez-Trincado}, Jos{\'e} G. and {Feuillet}, Diane and {Gallart}, Carme and {Hearty}, Fred R. and {Holtzman}, Jon and {Manchado}, Arturo and {Mu{\~n}oz}, Ricardo R. and {O'Connell}, Robert and {Rosado}, Margarita},
        title = "{The Lazy Giants: APOGEE Abundances Reveal Low Star Formation Efficiencies in the Magellanic Clouds}",
      journal = {\apj},
         year = 2020,
        month = jun,
       volume = {895},
       number = {2},
          eid = {88},
        pages = {88},
          doi = {10.3847/1538-4357/ab7305},
archivePrefix = {arXiv},
       eprint = {1901.03448},
 primaryClass = {astro-ph.GA},
       adsurl = {https://ui.adsabs.harvard.edu/abs/2020ApJ...895...88N}
}

@ARTICLE{YCQ2019,
       author = {{Yang}, Chengqun and {Xue}, Xiang-Xiang and {Li}, Jing and {Liu}, Chao and {Zhang}, Bo and {Rix}, Hans-Walter and {Zhang}, Lan and {Zhao}, Gang and {Tian}, Hao and {Zhong}, Jing and {Xing}, Qianfan and {Wu}, Yaqian and {Li}, Chengdong and {Carlin}, Jeffrey L. and {Chang}, Jiang},
        title = "{Tracing Kinematic and Chemical Properties of Sagittarius Stream by K-Giants, M-Giants, and BHB stars}",
      journal = {\apj},
         year = 2019,
        month = dec,
       volume = {886},
       number = {2},
          eid = {154},
        pages = {154},
          doi = {10.3847/1538-4357/ab48e2},
archivePrefix = {arXiv},
       eprint = {1909.12558},
 primaryClass = {astro-ph.GA},
       adsurl = {https://ui.adsabs.harvard.edu/abs/2019ApJ...886..154Y}
}

@ARTICLE{Cunningham2023,
       author = {{Cunningham}, Emily C. and {Hunt}, Jason A.~S. and {Price-Whelan}, Adrian M. and {Johnston}, Kathryn V. and {Ness}, Melissa K. and {Lu}, Yuxi and {Escala}, Ivanna and {Stelea}, Ioana A.},
        title = "{Chemical Cartography of the Sagittarius Stream with Gaia}",
      journal = {arXiv e-prints},
         year = 2023,
        month = jul,
          eid = {arXiv:2307.08730},
        pages = {arXiv:2307.08730},
          doi = {10.48550/arXiv.2307.08730},
archivePrefix = {arXiv},
       eprint = {2307.08730},
 primaryClass = {astro-ph.GA},
       adsurl = {https://ui.adsabs.harvard.edu/abs/2023arXiv230708730C}
}

@ARTICLE{Vasiliev2021,
       author = {{Vasiliev}, Eugene and {Belokurov}, Vasily and {Erkal}, Denis},
        title = "{Tango for three: Sagittarius, LMC, and the Milky Way}",
      journal = {\mnras},
         year = 2021,
        month = feb,
       volume = {501},
       number = {2},
        pages = {2279-2304},
          doi = {10.1093/mnras/staa3673},
archivePrefix = {arXiv},
       eprint = {2009.10726},
 primaryClass = {astro-ph.GA},
       adsurl = {https://ui.adsabs.harvard.edu/abs/2021MNRAS.501.2279V}
}

@ARTICLE{GC1,
       author = {{Goldsbury}, Ryan and {Richer}, Harvey B. and {Anderson}, Jay and {Dotter}, Aaron and {Sarajedini}, Ata and {Woodley}, Kristin},
        title = "{The ACS Survey of Galactic Globular Clusters. X. New Determinations of Centers for 65 Clusters}",
      journal = {\aj},
         year = 2010,
        month = dec,
       volume = {140},
       number = {6},
        pages = {1830-1837},
          doi = {10.1088/0004-6256/140/6/1830},
archivePrefix = {arXiv},
       eprint = {1008.2755},
 primaryClass = {astro-ph.GA},
       adsurl = {https://ui.adsabs.harvard.edu/abs/2010AJ....140.1830G}
}

@BOOK{GC2,
       author = {Sinnott, Roger W.},
        title = "{NGC 2000.0: The Complete New General Catalogue and Index Catalogues of Nebulae and Star Clusters by J. L. E. Dreyer}",
         year = 1988,
        publisher = "Cambridge University Press and Sky Publishing Corporation",
       adsurl = {https://ui.adsabs.harvard.edu/abs/1988cngc.book.....S}
}

@ARTICLE{GC3,
       author = {{Di Criscienzo}, M. and {Caputo}, F. and {Marconi}, M. and {Musella}, I.},
        title = "{RR Lyrae-based calibration of the Globular Cluster Luminosity Function}",
      journal = {\mnras},
         year = 2006,
        month = feb,
       volume = {365},
       number = {4},
        pages = {1357-1366},
          doi = {10.1111/j.1365-2966.2005.09819.x},
archivePrefix = {arXiv},
       eprint = {astro-ph/0511128},
 primaryClass = {astro-ph},
       adsurl = {https://ui.adsabs.harvard.edu/abs/2006MNRAS.365.1357D}
}

@ARTICLE{GC4,
       author = {{Shao}, Zhengyi and {Li}, Lu},
        title = "{Gaia parallax of Milky Way globular clusters - A solution of mixture model}",
      journal = {\mnras},
         year = 2019,
        month = nov,
       volume = {489},
       number = {3},
        pages = {3093-3101},
          doi = {10.1093/mnras/stz2317},
archivePrefix = {arXiv},
       eprint = {1908.06031},
 primaryClass = {astro-ph.GA},
       adsurl = {https://ui.adsabs.harvard.edu/abs/2019MNRAS.489.3093S}
}

@ARTICLE{StarHorse,
       author = {{Queiroz}, A.~B.~A. and {Anders}, F. and {Chiappini}, C. and {Khalatyan}, A. and {Santiago}, B.~X. and {Steinmetz}, M. and {Valentini}, M. and {Miglio}, A. and {Bossini}, D. and {Barbuy}, B. and {Minchev}, I. and {Minniti}, D. and {Garc{\'\i}a Hern{\'a}ndez}, D.~A. and {Schultheis}, M. and {Beaton}, R.~L. and {Beers}, T.~C. and {Bizyaev}, D. and {Brownstein}, J.~R. and {Cunha}, K. and {Fern{\'a}ndez-Trincado}, J.~G. and {Frinchaboy}, P.~M. and {Lane}, R.~R. and {Majewski}, S.~R. and {Nataf}, D. and {Nitschelm}, C. and {Pan}, K. and {Roman-Lopes}, A. and {Sobeck}, J.~S. and {Stringfellow}, G. and {Zamora}, O.},
        title = "{From the bulge to the outer disc: StarHorse stellar parameters, distances, and extinctions for stars in APOGEE DR16 and other spectroscopic surveys}",
      journal = {\aap},
         year = 2020,
        month = jun,
       volume = {638},
          eid = {A76},
        pages = {A76},
          doi = {10.1051/0004-6361/201937364},
archivePrefix = {arXiv},
       eprint = {1912.09778},
 primaryClass = {astro-ph.GA},
       adsurl = {https://ui.adsabs.harvard.edu/abs/2020A&A...638A..76Q}
}

@ARTICLE{GSEWWB,
       author = {{Wu}, Wenbo and {Zhao}, Gang and {Xue}, Xiang-Xiang and {Bird}, Sarah A. and {Yang}, Chengqun},
        title = "{Contribution of Gaia Sausage to the Galactic Stellar Halo Revealed by K Giants and Blue Horizontal Branch Stars from the Large Sky Area Multi-Object Fiber Spectroscopic Telescope, Sloan Digital Sky Survey, and Gaia}",
      journal = {\apj},
         year = 2022,
        month = jan,
       volume = {924},
       number = {1},
          eid = {23},
        pages = {23},
          doi = {10.3847/1538-4357/ac31ac},
archivePrefix = {arXiv},
       eprint = {2110.15571},
 primaryClass = {astro-ph.GA},
       adsurl = {https://ui.adsabs.harvard.edu/abs/2022ApJ...924...23W}
}

@ARTICLE{Feuillet2020,
       author = {{Feuillet}, Diane K. and {Feltzing}, Sofia and {Sahlholdt}, Christian L. and {Casagrande}, Luca},
        title = "{The SkyMapper-Gaia RVS view of the Gaia-Enceladus-Sausage - an investigation of the metallicity and mass of the Milky Way's last major merger}",
      journal = {\mnras},
         year = 2020,
        month = sep,
       volume = {497},
       number = {1},
        pages = {109-124},
          doi = {10.1093/mnras/staa1888},
archivePrefix = {arXiv},
       eprint = {2003.11039},
 primaryClass = {astro-ph.GA},
       adsurl = {https://ui.adsabs.harvard.edu/abs/2020MNRAS.497..109F}
}

@ARTICLE{Koppelman2019T,
       author = {{Koppelman}, Helmer H. and {Helmi}, Amina and {Massari}, Davide and {Price-Whelan}, Adrian M. and {Starkenburg}, Tjitske K.},
        title = "{Multiple retrograde substructures in the Galactic halo: A shattered view of Galactic history}",
      journal = {\aap},
         year = 2019,
        month = nov,
       volume = {631},
          eid = {L9},
        pages = {L9},
          doi = {10.1051/0004-6361/201936738},
archivePrefix = {arXiv},
       eprint = {1909.08924},
 primaryClass = {astro-ph.GA},
       adsurl = {https://ui.adsabs.harvard.edu/abs/2019A&A...631L...9K}
}

@ARTICLE{HS2019,
       author = {{Koppelman}, Helmer H. and {Helmi}, Amina and {Massari}, Davide and {Roelenga}, Sebastian and {Bastian}, Ulrich},
        title = "{Characterization and history of the Helmi streams with Gaia DR2}",
      journal = {\aap},
         year = 2019,
        month = may,
       volume = {625},
          eid = {A5},
        pages = {A5},
          doi = {10.1051/0004-6361/201834769},
archivePrefix = {arXiv},
       eprint = {1812.00846},
 primaryClass = {astro-ph.GA},
       adsurl = {https://ui.adsabs.harvard.edu/abs/2019A&A...625A...5K}
}

@ARTICLE{Horta2023,
       author = {{Horta}, Danny and {Schiavon}, Ricardo P. and {Mackereth}, J. Ted and {Weinberg}, David H. and {Hasselquist}, Sten and {Feuillet}, Diane and {O'Connell}, Robert W. and {Anguiano}, Borja and {Allende-Prieto}, Carlos and {Beaton}, Rachael L. and {Bizyaev}, Dmitry and {Cunha}, Katia and {Geisler}, Doug and {Garc{\'\i}a-Hern{\'a}ndez}, D.~A. and {Holtzman}, Jon and {J{\"o}nsson}, Henrik and {Lane}, Richard R. and {Majewski}, Steve R. and {M{\'e}sz{\'a}ros}, Szabolcs and {Minniti}, Dante and {Nitschelm}, Christian and {Shetrone}, Matthew and {Smith}, Verne V. and {Zasowski}, Gail},
        title = "{The chemical characterization of halo substructure in the Milky Way based on APOGEE}",
      journal = {\mnras},
         year = 2023,
        month = apr,
       volume = {520},
       number = {4},
        pages = {5671-5711},
          doi = {10.1093/mnras/stac3179},
archivePrefix = {arXiv},
       eprint = {2204.04233},
 primaryClass = {astro-ph.GA},
       adsurl = {https://ui.adsabs.harvard.edu/abs/2023MNRAS.520.5671H}
}

@ARTICLE{Pontus,
       author = {{Malhan}, Khyati},
        title = "{A New Member of the Milky Way's Family Tree: Characterizing the Pontus Merger of Our Galaxy}",
      journal = {\apjl},
         year = 2022,
        month = may,
       volume = {930},
       number = {1},
          eid = {L9},
        pages = {L9},
          doi = {10.3847/2041-8213/ac67da},
archivePrefix = {arXiv},
       eprint = {2204.09058},
 primaryClass = {astro-ph.GA},
       adsurl = {https://ui.adsabs.harvard.edu/abs/2022ApJ...930L...9M}
}

@article{nissen2010two,
  title={Two distinct halo populations in the solar neighborhood-Evidence from stellar abundance ratios and kinematics},
  author={Nissen, Poul Erik and Schuster, William J},
  journal={Astronomy \& Astrophysics},
  volume={511},
  pages={L10},
  year={2010},
  publisher={EDP Sciences}
}

@article{kruijssen2020kraken,
  title={Kraken reveals itself--the merger history of the Milky Way reconstructed with the E-MOSAICS simulations},
  author={Kruijssen, JM Diederik and Pfeffer, Joel L and Chevance, M{\'e}lanie and Bonaca, Ana and Trujillo-Gomez, Sebastian and Bastian, Nate and Reina-Campos, Marta and Crain, Robert A and Hughes, Meghan E},
  journal={Monthly Notices of the Royal Astronomical Society},
  volume={498},
  number={2},
  pages={2472--2491},
  year={2020},
  publisher={Oxford University Press}
}

@ARTICLE{Sgrstream,
       author = {{Ramos}, P. and {Antoja}, T. and {Yuan}, Z. and {Arentsen}, A. and {Oria}, P. -A. and {Famaey}, B. and {Ibata}, R. and {Mateu}, C. and {Carballo-Bello}, J.~A.},
        title = "{The Sagittarius stream in Gaia Early Data Release 3 and the origin of the bifurcations}",
      journal = {\aap},
         year = 2022,
        month = oct,
       volume = {666},
          eid = {A64},
        pages = {A64},
          doi = {10.1051/0004-6361/202142830},
archivePrefix = {arXiv},
       eprint = {2112.02105},
 primaryClass = {astro-ph.GA},
       adsurl = {https://ui.adsabs.harvard.edu/abs/2022A&A...666A..64R}
}

@ARTICLE{TheGaiamission,
       author = {{Gaia Collaboration} and others},
        title = "{The Gaia mission}",
      journal = {\aap},
         year = 2016,
        month = nov,
       volume = {595},
          eid = {A1},
        pages = {A1},
          doi = {10.1051/0004-6361/201629272},
archivePrefix = {arXiv},
       eprint = {1609.04153},
 primaryClass = {astro-ph.IM},
       adsurl = {https://ui.adsabs.harvard.edu/abs/2016A&A...595A...1G}
}

@ARTICLE{Thetaleofthetail,
       author = {{Amarante}, Jo{\~a}o A.~S. and {Smith}, Martin C. and {Boeche}, Corrado},
        title = "{The tale of the tail - disentangling the high transverse velocity stars in Gaia DR2}",
      journal = {\mnras},
         year = 2020,
        month = jan,
       volume = {492},
       number = {3},
        pages = {3816-3828},
          doi = {10.1093/mnras/staa077},
archivePrefix = {arXiv},
       eprint = {1912.12679},
 primaryClass = {astro-ph.GA},
       adsurl = {https://ui.adsabs.harvard.edu/abs/2020MNRAS.492.3816A}
}

@ARTICLE{GaiaDR3,
       author = {{Gaia Collaboration} and others},
        title = "{Gaia Data Release 3. Summary of the content and survey properties}",
      journal = {\aap},
         year = 2023,
        month = jun,
       volume = {674},
          eid = {A1},
        pages = {A1},
          doi = {10.1051/0004-6361/202243940},
archivePrefix = {arXiv},
       eprint = {2208.00211},
 primaryClass = {astro-ph.GA},
       adsurl = {https://ui.adsabs.harvard.edu/abs/2023A&A...674A...1G}
}

@ARTICLE{GaiaDR1,
       author = {{Gaia Collaboration} and others},
        title = "{Gaia Data Release 1. Summary of the astrometric, photometric, and survey properties}",
      journal = {\aap},
         year = 2016,
        month = nov,
       volume = {595},
          eid = {A2},
        pages = {A2},
          doi = {10.1051/0004-6361/201629512},
archivePrefix = {arXiv},
       eprint = {1609.04172},
 primaryClass = {astro-ph.IM},
       adsurl = {https://ui.adsabs.harvard.edu/abs/2016A&A...595A...2G}
}

@article{rix2022poor,
  title={The poor old heart of the Milky Way},
  author={Rix, Hans-Walter and Chandra, Vedant and Andrae, Ren{\'e} and Price-Whelan, Adrian M and Weinberg, David H and Conroy, Charlie and Fouesneau, Morgan and Hogg, David W and De Angeli, Francesca and Naidu, Rohan P and others},
  journal={The Astrophysical Journal},
  volume={941},
  number={1},
  pages={45},
  year={2022},
  publisher={IOP Publishing}
}

@article{sanders2016review,
  title={A review of action estimation methods for galactic dynamics},
  author={Sanders, Jason L and Binney, James},
  journal={Monthly Notices of the Royal Astronomical Society},
  volume={457},
  number={2},
  pages={2107--2121},
  year={2016},
  publisher={Oxford University Press}
}

@article{amarante2022gastro,
  title={Gastro Library. I. The Simulated Chemodynamical Properties of Several Gaia--Sausage--Enceladus-like Stellar Halos},
  author={Amarante, Jo{\~a}o AS and Debattista, Victor P and Silva, Leandro Beraldo E and Laporte, Chervin FP and Deg, Nathan},
  journal={The Astrophysical Journal},
  volume={937},
  number={1},
  pages={12},
  year={2022},
  publisher={IOP Publishing}
}

@article{ahumada202016th,
  title={The 16th data release of the sloan digital sky surveys: first release from the APOGEE-2 southern survey and full release of eBOSS spectra},
  author={Ahumada, Romina and Prieto, Carlos Allende and Almeida, Andr{\'e}s and Anders, Friedrich and Anderson, Scott F and Andrews, Brett H and Anguiano, Borja and Arcodia, Riccardo and Armengaud, Eric and Aubert, Marie and others},
  journal={The Astrophysical Journal Supplement Series},
  volume={249},
  number={1},
  pages={3},
  year={2020},
  publisher={IOP Publishing}
}

@article{zhao2015halo,
  title={Halo stream candidates in the LAMOST DR2},
  author={Zhao, Jing-Kun and Zhao, Gang and Chen, Yu-Qin and Tan, Ke-Feng and Gao, Meng-Tian and Yang, Ming and Zhang, Yong and Hou, Yong-Hui},
  journal={Research in Astronomy and Astrophysics},
  volume={15},
  number={8},
  pages={1378},
  year={2015},
  publisher={IOP Publishing}
}

@article{zhao2021low,
  title={Low-$\alpha$ metal-rich stars with sausage kinematics in the LAMOST survey: Are they from the Gaia-Sausage-Enceladus galaxy?},
  author={Zhao, Gang and Chen, Yuqin},
  journal={Science China Physics, Mechanics \& Astronomy},
  volume={64},
  number={3},
  pages={239562},
  year={2021},
  publisher={Springer}
}

@article{yan2018nature,
  title={The nature of the lithium enrichment in the most Li-rich giant star},
  author={Yan, Hong-Liang and Shi, Jian-Rong and Zhou, Yu-Tao and Chen, Yong-Shou and Li, Er-Tao and Zhang, Suyalatu and Bi, Shao-Lan and Wu, Ya-Qian and Li, Zhi-Hong and Guo, Bing and others},
  journal={Nature Astronomy},
  volume={2},
  number={10},
  pages={790--795},
  year={2018},
  publisher={Nature Publishing Group UK London}
}

@article{xing2023metal,
  title={A metal-poor star with abundances from a pair-instability supernova},
  author={Xing, Qian-Fan and Zhao, Gang and Liu, Zheng-Wei and Heger, Alexander and Han, Zhan-Wen and Aoki, Wako and Chen, Yu-Qin and Ishigaki, Miho N and Li, Hai-Ning and Zhao, Jing-Kun},
  journal={Nature},
  pages={1--4},
  year={2023},
  publisher={Nature Publishing Group UK London}
}

@article{xing2019evidence,
  title={Evidence for the accretion origin of halo stars with an extreme r-process enhancement},
  author={Xing, Qian-Fan and Zhao, Gang and Aoki, Wako and Honda, Satoshi and Li, Hai-Ning and Ishigaki, Miho N and Matsuno, Tadafumi},
  journal={Nature Astronomy},
  volume={3},
  number={7},
  pages={631--635},
  year={2019},
  publisher={Nature Publishing Group UK London}
}

@article{li2018catalog,
  title={A catalog of 10,000 very metal-poor stars from LAMOST DR3},
  author={Li, Haining and Tan, Kefeng and Zhao, Gang},
  journal={The Astrophysical Journal Supplement Series},
  volume={238},
  number={2},
  pages={16},
  year={2018},
  publisher={IOP Publishing}
}

@article{yan2021most,
  title={Most lithium-rich low-mass evolved stars revealed as red clump stars by asteroseismology and spectroscopy},
  author={Yan, Hong-Liang and Zhou, Yu-Tao and Zhang, Xianfei and Li, Yaguang and Gao, Qi and Shi, Jian-Rong and Zhao, Gang and Aoki, Wako and Matsuno, Tadafumi and Li, Yan and others},
  journal={Nature Astronomy},
  volume={5},
  number={1},
  pages={86--93},
  year={2021},
  publisher={Nature Publishing Group UK London}
}

@article{bland2016galaxy,
  title={The galaxy in context: structural, kinematic, and integrated properties},
  author={Bland-Hawthorn, Joss and Gerhard, Ortwin},
  journal={Annual Review of Astronomy and Astrophysics},
  volume={54},
  pages={529--596},
  year={2016},
  publisher={Annual Reviews}
}

@ARTICLE{Belokurov2022Aurora,
       author = {{Belokurov}, Vasily and {Kravtsov}, Andrey},
        title = "{From dawn till disc: Milky Way's turbulent youth revealed by the APOGEE+Gaia data}",
      journal = {\mnras},
         year = 2022,
        month = jul,
       volume = {514},
       number = {1},
        pages = {689-714},
          doi = {10.1093/mnras/stac1267},
archivePrefix = {arXiv},
       eprint = {2203.04980},
 primaryClass = {astro-ph.GA},
       adsurl = {https://ui.adsabs.harvard.edu/abs/2022MNRAS.514..689B}
}

@article{fernandez2019chemodynamics,
  title={Chemodynamics of newly identified giants with a globular cluster like abundance patterns in the bulge, disc, and halo of the Milky Way},
  author={Fern{\'a}ndez-Trincado, Jos{\'e} G and Beers, Timothy C and Tang, Baitian and Moreno, Edmundo and P{\'e}rez-Villegas, Angeles and Ortigoza-Urdaneta, Mario},
  journal={Monthly Notices of the Royal Astronomical Society},
  volume={488},
  number={2},
  pages={2864--2880},
  year={2019},
  publisher={Oxford University Press}
}

@article{fernandez2022galactic,
  title={Galactic ArchaeoLogIcaL ExcavatiOns (GALILEO)-I. An updated census of APOGEE N-rich giants across the Milky Way},
  author={Fern{\'a}ndez-Trincado, Jos{\'e} G and Beers, Timothy C and Barbuy, Beatriz and Minniti, Dante and Chiappini, Cristina and Garro, Elisa R and Tang, Baitian and Alves-Brito, Alan and Villanova, Sandro and Geisler, Doug and others},
  journal={Astronomy \& Astrophysics},
  volume={663},
  pages={A126},
  year={2022},
  publisher={EDP Sciences}
}

@article{ortigoza2023galactic,
  title={Galactic ArchaeoLogIcaL ExcavatiOns (GALILEO) II. t-SNE Portrait of Local Fossil Relics and Structures},
  author={Ortigoza-Urdaneta, Mario and Vieira, Katherine and Fern{\'a}ndez-Trincado, Jos{\'e} G and Queiroz, Anna and Barbuy, Beatriz and Beers, Timothy C and Chiappini, Cristina and Anders, Friedrich and Minniti, Dante and Tang, Baitian and others},
  journal={arXiv preprint arXiv:2306.08677},
  year={2023}
}

@article{anders2022photo,
  title={Photo-astrometric distances, extinctions, and astrophysical parameters for Gaia EDR3 stars brighter than G= 18.5},
  author={Anders, Friedrich and Khalatyan, A and Queiroz, Anna B{\'a}rbara de Andrade and Chiappini, Cristina and Ard{\`e}vol, J and Casamiquela, Laia and Figueras, F and Jim{\'e}nez-Arranz, {\'O} and Jordi, Carme and Mongui{\'o}, M and others},
  journal={Astronomy \& Astrophysics},
  volume={658},
  pages={A91},
  year={2022},
  publisher={EDP Sciences}
}

@article{van2008visualizing,
  title={Visualizing data using t-SNE.},
  author={Van der Maaten, Laurens and Hinton, Geoffrey},
  journal={Journal of machine learning research},
  volume={9},
  number={11},
  year={2008}
}

@article{lindegren2021gaia,
  title={Gaia Early Data Release 3-Parallax bias versus magnitude, colour, and position},
  author={Lindegren, L and Bastian, U and Biermann, M and Bombrun, A and De Torres, A and Gerlach, E and Geyer, R and Hern{\'a}ndez, J and Hilger, T and Hobbs, D and others},
  journal={Astronomy \& Astrophysics},
  volume={649},
  pages={A4},
  year={2021},
  publisher={EDP Sciences}
}

@article{hasselquist2021apogee,
  title={APOGEE chemical abundance patterns of the massive Milky Way satellites},
  author={Hasselquist, Sten and Hayes, Christian R and Lian, Jianhui and Weinberg, David H and Zasowski, Gail and Horta, Danny and Beaton, Rachael and Feuillet, Diane K and Garro, Elisa R and Gallart, Carme and others},
  journal={The Astrophysical Journal},
  volume={923},
  number={2},
  pages={172},
  year={2021},
  publisher={IOP Publishing}
}

@article{holtzman2015abundances,
  title={Abundances, stellar parameters, and spectra from the sdss-iii/apogee survey},
  author={Holtzman, Jon A and Shetrone, Matthew and Johnson, Jennifer A and Prieto, Carlos Allende and Anders, Friedrich and Andrews, Brett and Beers, Timothy C and Bizyaev, Dmitry and Blanton, Michael R and Bovy, Jo and others},
  journal={The Astronomical Journal},
  volume={150},
  number={5},
  pages={148},
  year={2015},
  publisher={IOP Publishing}
}

@ARTICLE{Bennett19,
       author = {{Bennett}, Morgan and {Bovy}, Jo},
        title = "{Vertical waves in the solar neighbourhood in Gaia DR2}",
      journal = {\mnras},
         year = 2019,
        month = jan,
       volume = {482},
       number = {1},
        pages = {1417-1425},
          doi = {10.1093/mnras/sty2813},
archivePrefix = {arXiv},
       eprint = {1809.03507},
 primaryClass = {astro-ph.GA},
       adsurl = {https://ui.adsabs.harvard.edu/abs/2019MNRAS.482.1417B}
}

@article{belokurov2023situ,
  title={In-situ vs accreted Milky Way globular clusters: a new classification method and implications for cluster formation},
  author={Belokurov, Vasily and Kravtsov, Andrey},
  journal={arXiv preprint arXiv:2309.15902},
  year={2023}
}

@ARTICLE{Limberg2023WK,
       author = {{Limberg}, Guilherme and {Ji}, Alexander P. and {Naidu}, Rohan P. and {Chiti}, Anirudh and {Rossi}, Silvia and {Usman}, Sam A. and {Ting}, Yuan-Sen and {Zaritsky}, Dennis and {Bonaca}, Ana and {Borbolato}, Lais and {Speagle}, Joshua S. and {Chandra}, Vedant and {Conroy}, Charlie},
        title = "{Extending the Chemical Reach of the H3 Survey: Detailed Abundances of the Dwarf-galaxy Stellar Stream Wukong/LMS-1}",
      journal = {arXiv e-prints},
         year = 2023,
        month = aug,
          eid = {arXiv:2308.13702},
        pages = {arXiv:2308.13702},
          doi = {10.48550/arXiv.2308.13702},
archivePrefix = {arXiv},
       eprint = {2308.13702},
 primaryClass = {astro-ph.GA},
       adsurl = {https://ui.adsabs.harvard.edu/abs/2023arXiv230813702L}
}

@article{an2021blueprint,
  title={A Blueprint for the Milky Way’s Stellar Populations. III. Spatial Distributions and Population Fractions of Local Halo Stars},
  author={An, Deokkeun and Beers, Timothy C},
  journal={The Astrophysical Journal},
  volume={918},
  number={2},
  pages={74},
  year={2021},
  publisher={IOP Publishing}
}

@article{schlafly2016optical,
  title={The optical--infrared extinction curve and its variation in the Milky Way},
  author={Schlafly, EF and Meisner, AM and Stutz, AM and Kainulainen, J and Peek, JEG and Tchernyshyov, K and Rix, H-W and Finkbeiner, DP and Covey, KR and Green, GM and others},
  journal={The Astrophysical Journal},
  volume={821},
  number={2},
  pages={78},
  year={2016},
  publisher={IOP Publishing}
}

@article{queiroz2018starhorse,
  title={StarHorse: a Bayesian tool for determining stellar masses, ages, distances, and extinctions for field stars},
  author={Queiroz, Anna B{\'a}rbara de Andrade and Anders, Friedrich and Santiago, Basilio Xavier and Chiappini, Cristina and Steinmetz, Matthias and Dal Ponte, Marina and Stassun, Keivan G and da Costa, Luiz N and Maia, Marcio Antonio Geimba and Crestani, J and others},
  journal={Monthly Notices of the Royal Astronomical Society},
  volume={476},
  number={2},
  pages={2556--2583},
  year={2018},
  publisher={Oxford University Press}
}

@article{naidu2021reconstructing,
  title={Reconstructing the Last Major Merger of the Milky Way with the H3 Survey},
  author={Naidu, Rohan P and Conroy, Charlie and Bonaca, Ana and Zaritsky, Dennis and Weinberger, Rainer and Ting, Yuan-Sen and Caldwell, Nelson and Tacchella, Sandro and Han, Jiwon Jesse and Speagle, Joshua S and others},
  journal={The Astrophysical Journal},
  volume={923},
  number={1},
  pages={92},
  year={2021},
  publisher={IOP Publishing}
}

\appendix
\section{Selection criteria}

\begin{table*}
	\centering
	\caption{Summary of the selection criteria we adopted to select dwarf galaxies and the literature we consulted.}
	\label{tab:DGs}
	\begin{tabular*}{\textwidth}{@{\extracolsep{\fill}}lccc}
		\hline
		Structures & Position & Criteria & References\\
		\hline
		LMC & $\rm \alpha = 80^\circ.893\;860$, $\rm \delta = -69^\circ.756\;126$ & 
            $d_{\rm{proj}}\leqq12^\circ$; $\rm 160\leqq RV\leqq348\;(km\;s^{-1})$; & (1)(2) \\ 
         & & $\rm{1.2\leqq PM_{\alpha}\leqq2.5\;(mas\;yr^{-1})}$; $\rm{-0.7\leqq PM_{\delta}\leqq1.5\;(mas\;yr^{-1})}$ & \\
        SMC &  $\rm \alpha = 13^\circ.186\;87$, $\rm \delta = -72^\circ.8286$ & 
            $d_{\rm{proj}}\leqq8^\circ$; $\rm 71\leqq RV\leqq 220\;(km\;s^{-1})$; & (1)(2) \\ 
         & & $\rm{0.2\leqq PM_{\alpha}\leqq1.8\;(mas\;yr^{-1})}$; $\rm{-1.8\leqq PM_{\delta}\leqq-0.8\;(mas\;yr^{-1})}$ & \\
        Sgr core &  $\Lambda \approx 0^\circ $, $B \rm{\approx 1^\circ.5} $ & 
            $|B|<10^\circ$; $b<0^\circ$; $\rm 90\leqq RV\leqq 220\;(km\;s^{-1})$; & (3)(4) \\ 
         & & $\rm{-3.5\leqq PM_{\alpha}\leqq-2\;(mas\;yr^{-1})}$; $\rm{-2.5\leqq PM_{\delta}\leqq-0.5\;(mas\;yr^{-1})}$ & \\
        Sgr stream & & Cross-match with Sgr stream catalogues & (5)(6) \\ 
		\hline
        \multicolumn{4}{@{}p{\textwidth}@{}}{\textit{Notes:} $(\Lambda,B)$ are coordinates in the right-handed coordinate system with respect to the Sgr stream as in \citet{Cunningham2023}, $d_{\rm{proj}}$ represents the projected distance on the celestial sphere.}\\
        \multicolumn{4}{@{}p{\textwidth}@{}}{\textit{References:} (1) \citet{Nidever2020}; (2) \citet{fernandes2023comparative}; (3) \citet{Vasiliev2021}; (4) \citet{Cunningham2023}; (5) \citet{YCQ2019}; (6) \citet{Sgrstream}}\\
    \end{tabular*}    
\end{table*}

\begin{table*}
	\centering
	\caption{The table exhibits the properties of 20 globular clusters we selected, and the literature we referred to during the selection. In particular, the $\rm{PM_{\alpha}}$, $\rm{PM_{\delta}}$, and $\rm{RV}$ columns respectively indicate the minimum and maximum values of the proper motion and radial velocity of the samples obtained by the selection method introduced in Section~\ref{section:sec3.1}.}
	\label{tab:GCs}
	\begin{tabular*}{\textwidth}{@{\extracolsep{\fill}}lccccccc}
	\hline
		Name & N stars & Center  & Angular size  & $\rm{PM_{\alpha}}$ & $\rm{PM_{\delta}}$ & $\rm{RV}$ & References\\
         & & $\rm{(\alpha,\delta)}$ & (arcmin) & $\rm{(mas\;yr^{-1})}$ & $\rm{(mas\;yr^{-1})}$ & $\rm{(km\;s^{-1})}$ & \\
	\hline
        NGC 3201 & 143 & $\rm {10^h17^m36.82^s,-46^\circ24'44.9"}$ & 18.2 & $7.870\sim8.771$ & $-2.363\sim-1.438$ &
        $468.14\sim506.36$ & (1)(2)\\
        NGC 5904 & 127 & $\rm {15^h18^m33.22^s,+02^\circ04'51.7"}$ & 17.4 & $3.786\sim4.483$ & $-10.284\sim-9.503$ &
        $36.01\sim69.77$ & (1)(2)\\
        NGC 5272 & 86 & $\rm {13^h42^m11.62^s,+28^\circ22'38.2"}$ & 16.2 & $-0.458\sim0.176$ & $-2.894\sim-2.362$ &
        $-161.77\sim-131.13$ & (1)(2)\\
        NGC 6341 & 47 & $\rm {17^h17^m07.39^s,+43^\circ08'09.4"}$ & 11.2 & $-5.144\sim-4.646$ & $-0.914\sim-0.445$ &
        $-133.81\sim-112.20$ & (1)(2)\\
        NGC 2808 & 40 & $\rm {09^h12^m03.10^s,-64^\circ51'48.6"}$ & 13.8 & $0.787\sim1.271$ & $-0.172\sim0.618$ &
        $89.55\sim125.43$ & (1)(2)\\
        NGC 7089 & 40 & $\rm {21^h33^m27.02^s,-00^\circ49'23.7"}$ & 12.9 & $3.141\sim3.708$ & $-2.480\sim-1.906$ &
        $-17.50\sim10.02$ & (1)(2)\\
        NGC 288 & 39 & $\rm {00^h52^m45.24^s,-26^\circ34'57.4"}$ & 19.2 & $4.033\sim4.328$ & $-5.877\sim-5.592$ &
        $-51.54\sim-36.89$ & (1)(4)\\
        NGC 362 & 37 & $\rm {01^h03^m14.26^s,-70^\circ50'55.6"}$ & 12.9 & $6.426\sim6.963$ & $-2.944\sim-2.373$ &
        $209.43\sim234.15$ & (1)(2)\\
        NGC 1904 & 35 & $\rm {05^h24^m10.59^s,-24^\circ31'27.3"}$ & 8.7 & $2.292\sim2.623$ & $-1.704\sim-1.473$ &
        $200.14\sim215.42$ & (2)(3)\\
        NGC 6101 & 32 & $\rm {16^h25^m48.12^s,-72^\circ12'07.9"}$ & 10.7 & $1.655\sim1.867$ & $-0.360\sim-0.159$ &
        $358.13\sim373.06$ & (1)(2)\\
        NGC 1851 & 28 & $\rm {05^h14^m06.76^s,-40^\circ02'47.6"}$ & 11.0 & $2.053\sim2.273$ & $-0.886\sim-0.498$ &
        $309.92\sim326.88$ & (1)(2)\\
        NGC 6779 & 27 & $\rm {19^h16^m35.57^s,+30^\circ11'00.5"}$ & 7.1 & $-2.174\sim-1.854$ & $1.431\sim1.759$ &
        $-148.02\sim-125.18$ & (1)(2)\\
        NGC 4590 & 25 & $\rm {12^h39^m27.98^s,-26^\circ44'38.6"}$ & 12.0 & $-2.802\sim-2.582$ & $1.592\sim1.888$ &
        $-100.39\sim-87.51$ & (1)(2)\\
        NGC 5024 & 24 & $\rm {13^h12^m55.25^s,+18^\circ10'05.4"}$ & 12.6 & $-0.274\sim-0.029$ & $-1.468\sim-1.234$ &
        $-71.21\sim-52.03$ & (1)(2)\\
        NGC 5466 & 20 & $\rm {14^h05^m27.29^s,+28^\circ32'04.0"}$ & 11.0 & $-5.415\sim-5.227$ & $-0.866\sim-0.736$ &
        $104.73\sim112.14$ & (1)(2)\\
        NGC 5286 & 18 & $\rm {13^h46^m26.81^s,-51^\circ22'27.3"}$ & 9.1 & $-0.074\sim0.520$ & $-0.390\sim0.049$ &
        $51.69\sim72.63$ & (1)(2)\\
        NGC 6934 & 18 & $\rm {20^h34^m11.37^s,+07^\circ24'16.1"}$ & 5.9 & $-2.743\sim-2.547$ & $-4.880\sim-4.559$ &
        $-419.38\sim-399.28$ & (1)(2)\\
        NGC 1261 & 16 & $\rm {03^h12^m16.21^s,-55^\circ12'58.4"}$ & 6.9 & $1.538\sim1.721$ & $-2.210\sim-2.010$ &
        $60.74\sim81.34$ & (1)(2)\\
        NGC 2298 & 16 & $\rm {06^h48^m59.41^s,-36^\circ00'19.1"}$ & 9.6 & $3.184\sim3.406$ & $-2.253\sim-2.065$ &
        $141.01\sim158.16$ & (1)(4)\\
        NGC 6981 & 13 & $\rm {20^h53^m27.70^s,-12^\circ32'14.3"}$ &  5.9 & $-1.325\sim-1.175$ & $-3.427\sim-3.294$ &
        $-334.19\sim-322.34$ & (1)(2)\\
	\hline
        \multicolumn{8}{@{}p{\textwidth}@{}}{\textit{References:} (1) \citet{GC1}; (2) \citet{GC2}; (3) \citet{GC3}; (4) \citet{GC4}}\\
	\end{tabular*}
\end{table*}

\begin{table*}
	\centering
	\caption{Summary of the selection criteria we employed to identify the substructures and the literature we utilized for reference.}
	\label{tab:Substructures}
	\begin{tabular*}{\textwidth}{@{\extracolsep{\fill}}lccccccc}
        \hline
        Substructures & Criteria & References\\
        \hline
        GSE & $|L_{z}|<0.5\;(\times\;10^{3}\;\rm{kpc\;km\;s^{-1}})$; $30<\sqrt{J_{r}}<50\;(\rm{kpc^{1/2}\;km^{1/2}\;s^{-1/2}})$; $r_{\rm{apo}}<40\;(\rm{kpc})$ & (1)(2) \\ 
        Thamnos & $\eta<-0.4$; $-1.8<E<-1.6\;(\times\;\rm{10^{5}\;km^{2}\;s^{-2}})$ & (3)(4) \\
        Sequoia & $-0.65<\eta<-0.4$; $-1.35<E<-1.0\;(\times\;\rm{10^{5}\;km^{2}\;s^{-2}})$ & (3) \\
        Helmi streams$^a$ & $0.75<J_{\phi}<1.7\;(\times\;10^{3}\;\rm{kpc\;km\;s^{-1}})$;  $1.6<L_{\perp}<3.2\;(\times\;10^{3}\;\rm{kpc\;km\;s^{-1}})$ & (4)(5) \\
        Wukong$^b$ & $0<J_{\phi}<1.0\;(\times\;10^{3}\;\rm{kpc\;km\;s^{-1}})$; $-1.7<E<-1.15\;(\times\;\rm{10^{5}\;km^{2}\;s^{-2}})$; $L_{y}>2.0\;(\times\;10^{3}\;\rm{kpc\;km\;s^{-1}})$; & (4)(6) \\
         & $(J_z-J_r)/J_{\rm{tot}}>0.3$; $90^\circ<$ arccos $(L_z/L)<120^\circ$ & \\
        Pontus & $-1.72<E<-1.56\;(\times\;\rm{10^{5}\;km^{2}\;s^{-2}})$; $-470<J_{\phi}<5\;(\rm{kpc\;km\;s^{-1}})$; $245<J_{r}<725\;(\rm{kpc\;km\;s^{-1}})$; & (7)(8) \\
         & $115<J_{z}<545\;(\rm{kpc\;km\;s^{-1}})$; $390<L_{\perp}<865\;(\rm{kpc\;km\;s^{-1}})$; $0.5<e<0.8$; $1<r_{\rm{peri}}<3\;(\rm{kpc})$; & \\
         &  $8<r_{\rm{apo}}<13\;(\rm{kpc})$ & \\
        \hline
        \multicolumn{3}{@{}p{\textwidth}@{}}{$^a$ To avoid the inclusion of Sgr, we impose a restriction that $|B|>10^\circ$}\\
        \multicolumn{3}{@{}p{\textwidth}@{}}{$^b$ We further exclude the sky areas where the distance may be overestimated as mentioned in Section~\ref{section:sec2.2.1}, specifically, the region where -50$^\circ$ $<$ $l$ $<$ 50$^\circ$ and -10$^\circ$ $<$ $b$ $<$ 15$^\circ$}\\
        \multicolumn{3}{@{}p{\textwidth}@{}}{\textit{Notes:} Circularity is represented with $\eta$, $L_\perp = \sqrt{L_x^2+L_y^2}$, and $e$ stands for eccentricity.}\\
        \multicolumn{3}{@{}p{\textwidth}@{}}{\textit{References:} (1) \citet{Feuillet2020}; (2) \citet{GSEWWB}; (3) \citet{Koppelman2019T}; (4) \citet{naidu2020evidence}; (5) \citet{HS2019}; (6) \citet{Limberg2023WK}; (7) \citet{Horta2023}; (8) \citet{Pontus}}\\
    \end{tabular*}    
\end{table*}

\section{2D ex-situ fraction map}

\begin{figure*}
 \includegraphics[width=\textwidth]{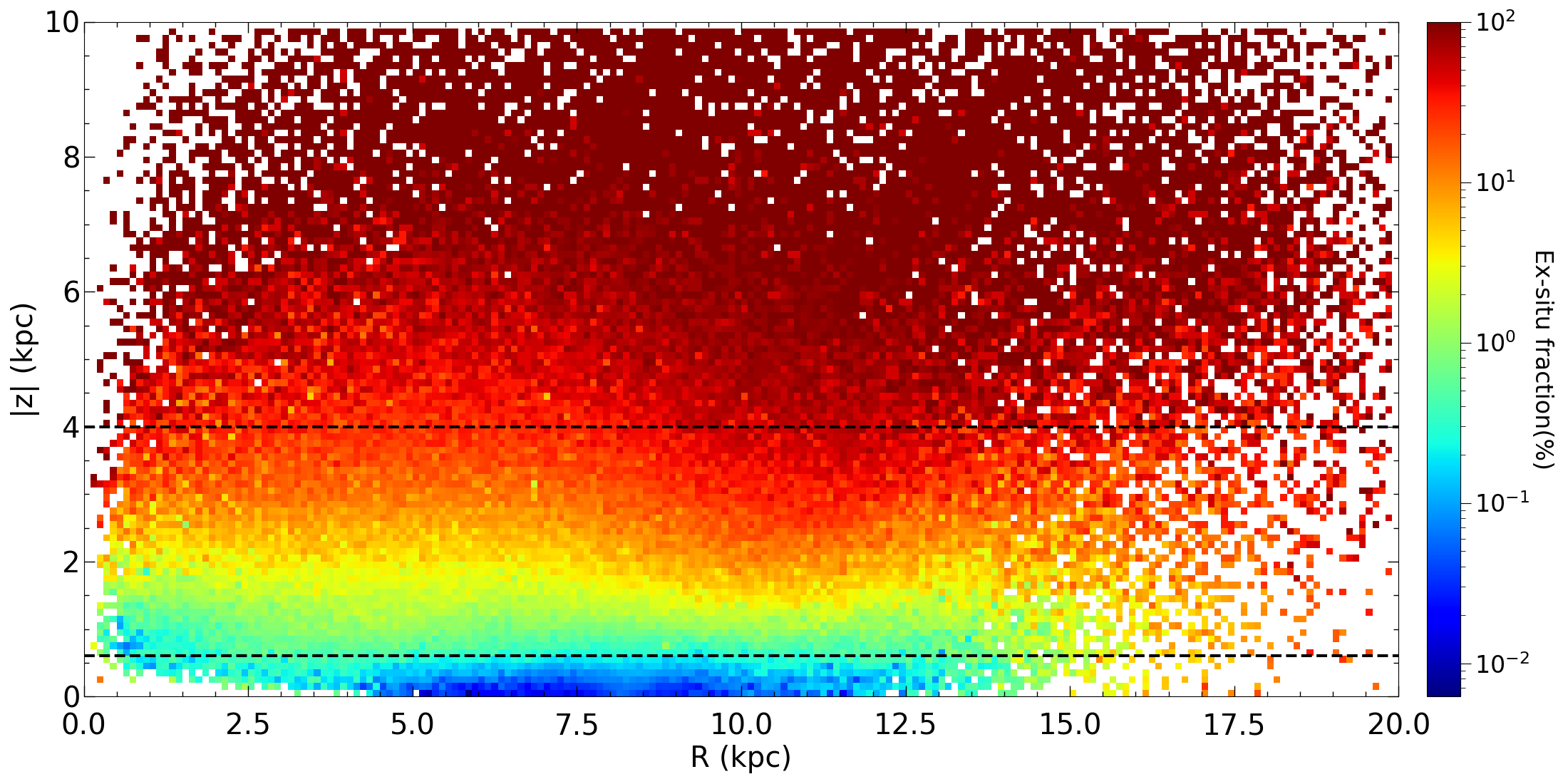}
 \caption{We illustrate the distribution of ex-situ fraction within the Galactocentric cylindrical coordinate system. The map is binned with a step size of 100 pc in both R and $|z|$ axis. The ex-situ fraction in each grid is visually represented through colour coding, which distinctly exhibits layering along the $|z|$-direction. According to the layering pattern, we divide our sample into three distinct regions: (1) a region characterized by an almost negligible presence of ex-situ stars ($|z|<$ 0.6 kpc); (2) a region where ex-situ stars are slightly interspersed (0.6 kpc $\leq|z|<$ 4 kpc); (3) a region predominantly occupied by ex-situ stars (4 kpc $\leq|z|\leq$ 10 kpc).}
 \label{fig:exsitu_percentage_2D}
\end{figure*}

\bsp	
\label{lastpage}
\end{document}